\documentclass[12pt,aj_pt4]{aastex}

\slugcomment{Submitted to Astronomical Journal}

\shorttitle{Stellar Abundances in WLM}
\shortauthors{Venn et al.}

\def\teff{T$_{\rm eff}$}
\def\logg{log g }
\def\etal{et\,al.\,}
\def\etall{et\,al.}
\def\kms{km\,s$^{-1}$}
\def\FeI{\ion{Fe}{1}}
\def\FeII{\ion{Fe}{2}}

\def\CrII{\ion{Cr}{2}}
\def\ScII{\ion{Sc}{2}}
\def\TiII{\ion{Ti}{2}}
\def\MgI{\ion{Mg}{1}}
\def\MgII{\ion{Mg}{2}}
\def\SiII{\ion{Si}{2}}
\def\SiIII{\ion{Si}{3}}
\def\HI{\ion{H}{1}}
\def\HII{\ion{H}{2}}

\def\OI{\ion{O}{1}}
\def\OII{\ion{O}{2}}
\def\OIII{{\ion{O}{3}}}
\def\NI{\ion{N}{1}}
\def\SrII{\ion{Sr}{2}}
\def\ZrII{\ion{Zr}{2}}
\def\BaII{\ion{Ba}{2}}
\def\NaI{\ion{Na}{1}}

\begin{document}


\title{The Chemical Composition of Two Supergiants in the 
       Dwarf Irregular Galaxy WLM}


\author{Kim A. Venn}
\affil{Macalester College, Saint Paul, MN, 55105; venn@macalester.edu \\
       University of Minnesota, 116 Church Street S.E., Minneapolis, MN, 55455}
%
\author{Eline Tolstoy}
\affil{Kapteyn Institute, University of Groningen, PO Box 800, 9700AV 
       Groningen, the Netherlands; etolstoy@astro.rug.nl}
\author{Andreas Kaufer}
\affil{European Southern Observatory, Alonso de Cordova 3107, 
       Santiago 19, Chile; akaufer@eso.org}
\author{Evan D. Skillman}
\affil{Department of Astronomy, University of Minnesota, 
       116 Church Street S.E., Minneapolis, MN, 55455;
       skillman@astro.umn.edu}
\author{Sonya M. Clarkson}
\affil{Macalester College, Saint Paul, MN, 55105}
\author{Stephen J. Smartt}
\affil{Institute of Astronomy, University of Cambridge, 
       Madingley Road, Cambridge, CB3 0HA, UK; sjs@ast.cam.ac.uk}
\author{Danny J. Lennon}
\affil{Isaac Newton Group of Telescopes (ING), Santa Cruz de La Palma, 
       Canary Islands, E-38700, Spain; djl@ing.iac.es}
\and
\author{Rolf P. Kudritzki}
\affil{Institute for Astronomy, University of Hawaii,
       2680 Woodlawn Drive, Honolulu, 95822; kud@ifa.hawaii.edu}

\begin{abstract}

The chemical composition of two stars in WLM have been determined
from high quality UVES data obtained at the VLT UT2\footnote{Based 
  on observations collected at the European 
  Southern Observatory, proposal number 65.N-0375}.
The model atmospheres analysis shows that they have the same metallicity, 
[Fe/H] = $-$0.38 $\pm$0.20 ({\it $\pm$0.29})\footnote{In this paper,
 we adopt the standard notation [$X$/H] = log($X$/H) - log($X$/H)$_\odot$.
 Also, abundances shall be reported with two uncertainties: the first is 
 the line-to-line scatter, and the second (in parentheses and italics) is
 an {\it estimate} of the systematic error due to uncertainties in the
 atmospheric parameters.}.    
Reliable magnesium abundances are determined from several lines of two
ionization states in both stars resulting in 
[Mg/Fe] = $-$0.24 $\pm$0.16 ({\it $\pm$0.28}).   
This result suggests that the [$\alpha$(Mg)/Fe] ratio in WLM may be
suppressed relative to solar abundances (also supported by
differential abundances relative to similar stars in 
NGC~6822 and the SMC).
The absolute Mg abundance, [Mg/H] = $-$0.62 is high relative
to what is expected from the nebulae though, where two independent 
spectroscopic analyses of the \HII\ regions in WLM yield [O/H] = $-$0.89.
Intriguingly, the oxygen abundance determined from the \OI\ $\lambda$6158
feature in one WLM star is [O/H] = $-$0.21 $\pm$0.10 ({\it $\pm$0.05}),
corresponding to five times higher than the nebular oxygen abundance. 
This is the first time that a significant difference between stellar 
and nebular oxygen abundances has been found, and presently, there is
no simple explanation for this difference.  The two stars are 
massive supergiants with distances that clearly place them in WLM.
They are young ($\le$10 Myr) and should have a similar composition to the ISM.
Additionally, differential abundances suggest that the O/Fe ratio in
the WLM star is consistent with similar stars in NGC6822 and the SMC,
galaxies where the average stellar oxygen abundances are in excellent 
agreement with the nebular results. 
If the stellar abundances reflect the true composition of WLM, 
then this galaxy lies well above the metallicity-luminosity 
relationship for dwarf irregular galaxies.  
It also suggests that WLM is more chemically evolved than currently 
interpreted from its color-magnitude diagram.
The similarities between the stars in WLM and NGC6822 suggest that these 
two galaxies may have had similar star formation histories. 

\end{abstract}

\keywords{galaxies: abundances, dwarf galaxies, individual (WLM) 
stars: abundances}

\section{Introduction \label{intro}}

The evolution of the chemical abundances in a galaxy is intimately 
linked to its star formation history (Tinsley 1979).   Different
elements are produced during the evolution of stars of different
masses, and over a range of timescales.    
If the star formation in a galaxy proceeds by a series of bursts, 
as suggested for dwarf galaxies (c.f., Matteucci \& Tosi 1985, 
and reviews by Hodge 1989 and Mateo 1998),
rather than smooth, approximately constant star formation that characterizes 
the evolution of galactic disks (e.g., Matteucci \& Greggio 1986, 
Edvardsson \etal 1993, Chiappini \etal 1997) 
then this should lead to clear differences 
in the evolution of the chemical abundances.   
One ratio of particular importance is the $\alpha$/Fe ratio.   Oxygen 
is produced primarily in high-mass stars of negligible lifetimes 
and ejected by SNe II, while iron is produced in both SNe II and SNe Ia.  
Stars that form shortly after the interstellar medium has been 
enriched by SNe II may have enriched $\alpha$/Fe 
ratios, while those that form sometime after the SNe Ia contribute 
will have lower $\alpha$/Fe ratios.    The timescale for changes in the 
$\alpha$/Fe ratio depends not only on the SFH, but also on the IMF, 
the SNe Ia timescale, and the timescales for mixing the SNe Ia and 
SNe II products back into the interstellar medium.
Gilmore and Wyse (1991) demonstrated the 
expected differences in $\alpha$/Fe ratios 
in galaxies with different star formation histories.   Most notably 
they write ``We emphasize that there is nothing special or universal 
about solar element ratios, and that one should not expect the solar 
neighborhood situation to be reproduced in any other environment which 
has had a different star formation history.''
The analysis of bright nebular emission lines of \HII\ regions
(and some PN) has been the most frequent approach to modeling
the chemical evolution of galaxies to date (see Pagel 1997),
and yet only a limited number of elements can be examined and
quantified when using this approach.  In particular, iron-group 
abundances, and thus the $\alpha$/Fe ratios, are either not possible
or carry significant uncertainties from nebular analyses alone.
Supergiant stars, however, have both $\alpha$ and iron-group element
absorption lines in their spectra, and it is possible to obtain a
reliable $\alpha$/Fe ratio from these stars\footnote{The most 
  reliable ratios, of particular $\alpha$ and Fe-group elements, varies 
  depending on the temperature of the supergiant (blue versus red) 
  analysed.  In the analysis of A-type supergiants, \OI\, \MgI\ and \MgII\
  usually provide the most reliable $\alpha$ element abundances, while 
  \CrII\ and \FeII\ provide the best Fe-group results.} 

Imaging of dwarf galaxies in the Local Group show a
wide variety of star formation histories (Grebel 1997, Mateo 1998). 
Measuring $\alpha$/Fe ratios in stars of various ages provides an
ideal way to test these star formation histories and constrain
the chemical evolution of these galaxies.    This is possible
for the closest galaxies from detailed analyses of their red 
giants, i.e., the Magellanic Clouds (Hill \etal 2003) and 
Galactic dSph galaxies (Bonifacio \etal 2000, 
Shetrone, Cote, \& Sargent 2001, 
Shetrone \etal 2003, Tolstoy \etal 2003). 
Some dwarf irregular galaxies appear to have undergone bursts of star
formation during their histories and thus should show a variety of
$\alpha$/Fe ratios throughout their evolution, but they are too distant
for detailed analysis of their red giant branch (=RGB, stars that can 
sample ages $>$1 Gyr).   
It is possible to observe and analyse the spectra of their bright,
young massive stars though,  since they continue to form stars today. 
These stars will reflect the {\it integrated} star formation 
history and chemical evolution in their $\alpha$/Fe abundances,
and yet Gilmore and Wyse (1991) predict that this
ratio could vary significantly from galaxy to galaxy.

While dwarf galaxies are the most common type of galaxy in the Local 
Group (c.f., van den Bergh 2000), the dwarf spheroidal 
galaxies are (almost) all located near the large mass galaxies, which 
suggests that their environment may be related to their morphology
and possibly star formation history (van den Bergh 1994).
On the other hand, dwarf irregular galaxies tend to be more 
isolated systems found on the outskirts of the Local Group.
Nebular analyses suggest that the low luminosity dwarf irregular galaxies 
are very metal-poor ($\sim$1/20 solar, Skillman \etal 1989a,b), suggesting
that they have undergone relatively little chemical evolution. 
These objects may be more similar to the protogalactic fragments 
at the time of Galaxy formation in hierarchical merging models;
low mass, gas rich, and metal-poor.

WLM (DDO221, UGCA444) is an isolated dwarf irregular galaxy and its
star formation history has been studied in detail with HST STIS imaging 
by Rejkuba \etal (2000) and HST WFPC2 imaging by Dolphin (2000), building
on the ground-based studies by Minniti \& Zijlstra (1997, hereafter MZ97) 
and Ferraro \etal (1989).    
Rejkuba \etall's detection of the horizontal branch confirmed 
the distance modulus of WLM at ($m-M$)$_0$ = 24.95 $\pm$0.13, the 
foreground reddening at E(V-I) = 0.03, and the presence of an ancient
population.  Their distance D = 0.98 $\pm$ 0.06 Mpc is 
within 1$\sigma$ of the D$_{WLM}$ = 0.95 summarized by van den Bergh 
(1994, 2000),
which includes a revision of the Cepheid distance originally found by 
Sandage \& Carlson (1985).  Dolphin used the CMD to examine the star 
formation and chemical enrichment history of WLM, finding that more
than half the stars in WLM formed over 9 Gyr ago.   The star formation 
rate has gradually decreased since then, with a recent increase that is 
concentrated in the bar of the galaxy.   
The metal enrichment appears to show a gradual increase from 
[Fe/H] $\le-2$ over 12 Gyr ago, reaching a current value near $-$1~dex.
This low present-day value is consistent with the published nebular 
abundance, 12+log(O/H) = 7.77 $\pm$0.17 based on three \HII\ regions 
(Hodge \& Miller 1995, hereafter HM95; Skillman \etal 1989a), or an abundance
relative to solar\footnote{Note that this ratio is based on a solar 
  oxygen abundance of 12 + log(O/H) = 8.66 from Asplund (2003, consistent
  with Allende-Prieto \etal 2001).}
of [O/H] = $-$0.89.
WLM lies well above the plane of the Local Group and is fairly
isolated, thus MZ97 argue, based on its distance to IC1613, 
that it could not have suffered a major encounter within the 
past Gyr to initiate its current star formation activity.
However, since then the Cetus dwarf spheroidal 
galaxy has been discovered (Whiting, Hau, \& Irwin 1999), which is 
located much closer to WLM than IC1613.  
Another possibility is if the one globular cluster in WLM 
(see also Hodge \etal 1999) passed near the center of the galaxy and 
caused a disruption, but the globular cluster space velocities are not known.  

The distance to WLM puts the tip of the red giant branch at
V $\sim$ 22, which is out of reach for detailed chemical 
abundance analyses of its RGB population.   However, there
are bright blue supergiants in WLM's central star forming 
region with V $\sim$ 18.   
These young, massive stars sample the current metallicity in the 
galaxy (thus the integrated metal enrichment over its lifetime),
and provide valuable end-point abundances, particularly
for the iron-group, as well as for the s-process; but they
do not provide information on the evolution of these abundances
(as in RGB analyses).
They can also provide a check on the accuracy of the $\alpha$
element abundances from nebular spectroscopy.    Model
atmospheres analyses of the massive stars in
Orion (Cunha \& Lambert 1994), the Large Magellanic Cloud 
(LMC; Hill 1999, Hill \etal 1995, Rolleston \etal 2002, Korn \etal 2002), 
the Small Magellanic Cloud 
(SMC; Venn 1999, Hill 1997, Rolleston \etal 2003), 
the Andromeda galaxy (M31; Venn \etal 2000, Trundle \etal 2002), 
and the dwarf irregular galaxy NGC6822 (Venn \etal 2001), 
have all determined stellar oxygen abundances in excellent agreement
with the nebular results.
In this paper, we present the first stellar abundances in WLM.

\section{Observations \& Reductions}

Spectra for two A-type supergiants and one B-type supergiant in WLM 
were taken at the VLT-UT2 with UVES (Dekker \etal 2000) between 
August 20$-$22, 2000 (Table~\ref{obs}).     
60-minute exposures of each star were made in (mostly) sub-arcsecond  
seeing conditions through a 1.0-arcsec slit.  On-chip binning (2x2) 
was used at readout, yielding R$\sim$32,000, or R$\sim$20,000 per
3 pixel resolution element.
A combined S/N ratio $\sim$30 per pixel, or S/N $\sim$50
per resolution element, was attained after coaddition.
Two dichroic settings (390/564 and non-standard 390/840) were 
used with standard calibrations and pipeline reduced.  
The pipeline reduction (Balester \etal 2000) includes bias
and interorder background subtraction, flatfield correction,
optimal extraction (with cosmic ray rejection) above the sky
level, and the wavelength calibration.
The wavelength range spanned 3800 \AA\,  
$\le$ W$_\lambda$ $\le$ 10250 \AA, with
gaps near 4600\,\AA, 5600\,\AA, and from 6650 - 8470\,\AA. 

Targets were selected from photometric BVI colors (from archival NTT imaging) 
and from low resolution spectroscopy (from WHT-ISIS and ESO 3.6m EFOSC
spectroscopy by D.J. Lennon and S.J. Smartt, and FORS2 spectroscopy 
by T. Szeifert).   We selected the brightest (V$<$18.5), apparently 
isolated stars, with colors ranging from $-$0.3 $\le$ (B-V)$_o$ $\le$ 0.6.
The six brightest stars from the original target list are in 
Table~\ref{sample}.
Unfortunately, further analysis of the UVES spectrum of the early 
B-type star WLM-35 is not possible because of missing \SiIII\ lines, 
needed for the effective temperature determination,
in the dichroic gap near 4600\,\AA.
The locations of these stars in WLM can be seen in Fig.~1
(coordinates available in Sandage \& Carlson 1985).

The two stars analysed here are WLM-15 and WLM-31.  
WLM-15 proved to be that of a normal, isolated A5 Ib
supergiant, with a radial velocity consistent with
WLM membership (Table~\ref{sample}).
WLM-31 is a blend of a normal mid-A supergiant and a 
foreground red giant.   The red giant spectrum has been 
subtracted for this analysis (discussed further below). 
This has been possible partially because it contributes 
very little light in the blue compared to the A-type 
supergiant and because its radial velocity, $\sim$ 0 \kms\
(which is consistent with the Galaxy halo stars at WLM's
Galactic coordinates), is well offset from the A-supergiant in WLM
($-$117 \kms).

Finally, we mention that spectra of hot, rapidly-rotating stars were 
taken as telluric divisors, but not used.   The S/N in our WLM spectra 
was too low in the red spectral regions where the telluric divisors would 
have been useful (e.g., \NI\ 8200 lines) for further analysis.   
There were no telluric lines near the \MgII\ 7880 features used in this 
analysis.

\section{Atmospheric Analyses \label{atmanal}}

The WLM A-supergiants have been analysed using ATLAS9 
(hydrostatic, line-blanketed, plane parallel) model atmospheres
(Kurucz 1979, 1888).   These atmospheres have been shown to be
appropriate for the photospheric analysis of lower luminosity
A-supergiants (Przybilla 2002), and have been successfully 
used for the photospheric analysis of stars in the Galaxy,
the Magellanic Clouds, and M31 
(Venn 1995a,b, 1999; Luck \etal 1998; Venn \etal 2000; 
Przybilla 2002).   
A-type supergiants require a tailored analysis
where only weak spectral lines (that typically form deep in
the photosphere) are included.   In this analysis, weak lines are 
defined as those where a change in microturbulence 
($\xi$), $\Delta\xi$ = $\pm$1 \kms, yields a change in abundance 
of log(X/H) $\le$ 0.1.   Typically, this was found to be 
W$_\lambda \le$ 160 m\AA.
Using weak lines exclusively also helps to avoid uncertainties
in the model atmospheres analysis due to neglected NLTE and
spherical extension effects in the atmospheric structure,
as well as NLTE and $\xi$ effects in the line formation calculations.
 
The critical spectral features used to determine the model
atmosphere parameters (effective temperature, \teff, and
gravity) are the wings of the H$\gamma$ line (see Fig.~2)
and ionization equilibrium of \MgI\ and \MgII\ (discussed below
for each star individually).
A locus of \teff-gravity pairs that reproduce the Balmer
line and Mg ionization equilibrium are shown
in Fig.~3.   
NLTE calculations are included for Mg using the model
atom developed by Gigas (1988) and a system of programs
first developed by W. Steenbock at Kiel University and
further developed and upgraded by M. Lemke (now associated 
with the Dr. Remeis-Sternwarte, Bamberg).  The results
from this model are in very good agreement with the more
recent Mg model developed by Przybilla \etal (2001a).  The
NLTE corrections for Mg are quite small;  for both stars, 
the \MgII\ corrections are $\le$0.03~dex, and for \MgI\ they 
average $-$0.11~dex (with a range from $-$0.02 to $-$0.16~dex).
Ionization equilibrium of \FeI\ and \FeII\ is also shown in 
Fig.~3, but only used as a check since NLTE effects may 
be important for \FeI\ (e.g., Boyarchuk \etal 1985, Gigas 1986).  
\FeII\ NLTE effects are negligible (Becker 1998).
 
Microturbulence has been found by examining the line 
abundances of \FeII, \TiII, \CrII\ (and \FeI), and require
no relationship with equivalent width.   The results
from \FeI\ were consistently lower, although  allowing for an
uncertainty in $\Delta\xi$ of $\pm$1\,\kms\ brings the results
from the different species into excellent agreement.
Considering the weak line nature of this analysis, 
a single value for $\xi$ was adopted for each star.

The atmospheric parameters determined for both stars are listed in 
Table~\ref{atms}.   The uncertainties in \teff\ for WLM-15, 
$\Delta$\teff = $\pm$200~K, is estimated from the range where 
log(\MgII) = log(\MgI) $\pm$0.2 when holding gravity fixed.   This 
range allows for uncertainties in equivalent width measurements, 
atomic data, and uncertainties in the NLTE calculations.    The 
uncertainty in \teff\ for WLM-31 is larger, $\Delta$\teff = $\pm$300~K,
because only one weak line of \MgII\ is available for the analysis
(we consider a signficant uncertainty on its equivalent width measurement
W$_\lambda$ = 25 $\pm$ 10 m\AA), 
and the measurements of the \MgI\ $\lambda$5180 lines are less certain 
(see the discussion on recovering the spectrum of WLM-31 from
a blend with a foreground red giant, Section ~\ref{recovery}).
The uncertainties in gravity, $\Delta$\logg=$\pm$0.1,  are estimated 
from the range in the H$\gamma$ profile fits while holding \teff\ 
fixed for both stars.

\subsection{Recovering the Intrinsic Spectrum of WLM 31 \label{recovery} }

WLM-31 shows sharp and narrow \MgI\ 5180 lines at the radial velocity offset 
of the WLM galaxy, but also strong and broad \MgI\,b lines at the Galactic 
rest wavelengths.  It appears to be an A-supergiant in WLM, combined with 
a foreground red giant.    

There are no HST images for WLM-31.
Examination of the FORS images in B, V, and I of WLM (with
seeing estimates of 0.75, 0.81, 1.07 arcseconds, respectively)
reveal that this object is sharp in the I image, yet in the 
B (and possibly V ) image, where the seeing estimate is 
the lowest and the color difference between the two stars would 
be the highest, there is a marginal asymmetry to the northeast.
A $\sim$40\% contribution by the RGB star to the V-band
continuum in the UVES spectrum of WLM-31 (discussed in the next
section), corresponds to a difference in magnitude of 
$\Delta$V $\le$ 0.5, within $\sim$0.7$''$.

Since the radial velocities and spectral types of these two 
stars are so different, we have attempted to recover and analyse 
the A-supergiant spectrum in WLM;  see Fig.~4.
As a first step, the normalized UVES spectrum of WLM-15 was combined
with of a variety of red giants (RGB UVES spectra from Shetrone \etal 2003).
Combining WLM-15 with a red giant near 4100~K (e.g., Fornax-M12) reproduces 
the 5200~\AA\ spectral region quite well if the RGB spectrum 
is weighted at 40\%.    The same was true near 4900~\AA\ if the RGB 
spectrum was weighted at 25\%.    
As a second step, the surface flux for a normal A-type supergiant 
(WLM-15 parameters) was compared to that of a metal-poor red giant
([Fe/H] = $-$1.5, \teff=4250~K, \logg=1.0) and a metal-poor red dwarf 
([Fe/H]=$-$1.5, \teff=4250~K, \logg=4.0) using Kurucz (1993) 
models\footnote{
  Available from stsci.edu at ftp.stsci.edu/cdbs/cdbs2/grid/k93models/}. 
If the red giant flux contributes 25\% near 4900~\AA, then the 
Kurucz fluxes reproduce our estimated flux ratio at 5200~\AA\ 
(as 37\%).   This estimate is the same for the red dwarf 
(since the continuum level is dominated by temperature).  
The Kurucz fluxes can also be used to estimate 
the contribution from the RGB stars to the A-supergiant's 
continuum below 4900~\AA.

The line list for WLM-31 was taken from our analysis of WLM-15
(since their spectra are quite similar).  Equivalent widths were 
measured, then scaled depending on their wavelength;  contamination 
by the RGB star has been estimated per wavelength by comparing 
Kurucz (1993) model atmospheres (as described above).   The scaling 
used for the equivalent widths was 37\%, 25\%, 20\%, 15\%, 9\%, and 2\% 
at 5200~\AA\, 4900~\AA, 4600~\AA, 4400~\AA\, 4200~\AA, and 4000~\AA,
respectively, based on the Kurucz fluxes, with linear interpolation 
used for wavelengths between these points. 
This scaling law varies slightly for Kurucz models with 
$\Delta$\teff = $\pm$250~K but does not affect our WLM-31 
elemental abundances ($\Delta$log($X$/H) $\le$ 0.02~dex).
Spectral lines above 5200~\AA\ were not used since the RGB star begins to 
dominate the spectrum, washing out the weak spectral lines in the A-supergiant
and making the continuum corrections more uncertain. 
When the abundances were calculated, 
the results from blue versus red absorption lines were in excellent 
agreement after this scaling, suggesting that  we have sufficiently
recovered the A-supergiant spectrum.
However, several lines observed in WLM-15 were not recovered in WLM-31. 
Very weak lines were not recovered because the S/N for the WLM-31 
spectrum is slightly lower, and some stronger lines were not recovered 
because WLM-31 is slightly cooler such that changing $\xi$ by 
1~\kms\ changed log($X$/H) by $\ge$0.1~dex (the definition for a
weak line to be used in this analysis). 
As a final step, each individual absorption line used in the WLM-31 
analysis was reviewed, and any lines that were clearly contaminated by 
a strong RGB line were discarded 
(of course, the RGB contaminating lines were from 
different elements and transitions because of the radial velocity offset). 
Thus, the set of line measurements listed in Table~\ref{lines} 
are considered to be the most reliable in the WLM-31 spectrum.

Clearly the abundances from WLM-31 should be viewed with
caution considering the difficulties deconvolving the two stars
(particularly not being able to clearly analyse the RGB contaminant 
itself because we do not know its actual atmospheric parameters).   
Nevertheless,
the spectral analysis of this star is remarkably consistent
between the blue and red spectral lines that we have carefully
selected in the spectrum, and from species to species (below).

\subsection {WLM Membership \label{membership}}

Because it is important to be as certain as possible that these 
stars are members of WLM, the atmospheric parameters determined 
above are used here to derive a spectroscopic distance.   Even 
though these distances are not very accurate, we do this as a 
consistency check and to ensure that they are not foreground
(post-AGB?) stars. 
The calculation is
straightforward for the isolated star, WLM-15, but difficult
for the blended star WLM-31 without knowing the difference in 
magnitude between the A-supergiant and the foreground red giant.

For WLM-15, we adopt 12 M$_\odot$ as a reasonable 
mass estimate from its atmospheric parameters and standard 
stellar evolution tracks (e.g., Lejeune \& Schaerer 2001). 
Low foreground reddening is consistently found to WLM 
(see discussion by Rejkuba \etal 2000).
Adopting E(B-V) = 0.03, and zero bolometric correction,
the distance to WLM-15 is $\sim$850~kpc.  
The mass estimate and gravity determination are
the most critical factors in this calculation; allowing 
mass to vary by $\pm$6 M$_\odot$ or gravity to vary by $\mp$0.2~dex
can change the distance by $\pm$200~kpc.   All other
parameters (e.g., temperature estimate, bolometric correction, and
internal reddening up to A$_{\rm v}$=0.4) have much smaller effects.
This stellar distance is in excellent agreement with that summarized
by van den Bergh (1994, 2000) of D$_{WLM}$ = 0.95~Mpc.

The distance to WLM-31 is much less certain because the A-supergiant
in WLM is blended with a foreground red giant, and we cannot
be as certain of its physical properties.  As an estimate, 
if the red giant contributes 40\% of the light in the V-band
(see Section 3.1),
then the magnitude of the remaining A-supergiant is $\sim$0.5 mag fainter,
or V $\sim$ 18.9.    Adopting the same mass and reddening used for WLM-15,
since they have similar atmospheric parameters, then the distance is
$\sim$1200~kpc.     This is in fair agreement with D$_{WLM}$. 
However, if WLM-31 is really 0.8 mag fainter than WLM-15, then either
WLM-31 must have higher (internal) foreground reddening or a lower mass.
For reddening, $\Delta$A$_v$ $\sim$0.7 is much higher than inferred from 
the CMD analyses; e.g., MZ97 set an upper limit to the differential 
reddening at 0.1 mag by comparing the RGB and blue main sequence on 
both sides of WLM.   However, Skillman \etal (1989) suggest an upper 
limit to the total extinction of A$_B$ $\le$0.4 to 0.5, and 
Hodge \& Miller (1995) find E(B-V) = 0.1 $\pm$0.1 for two \HII\ regions 
in WLM, or A$_v$ $\le$0.6.   These high values could bring 
WLM-31 into agreement with WLM-15 (which would need to have 
zero reddening then).  
On the other hand, if WLM-31 had a mass near 6 M$_\odot$, 
instead of 12 M$_\odot$ adopted for WLM-15, and given their 
similar atmospheric parameters, then this would imply a 
difference of $\sim$0.3 in luminosity, or $\sim$0.8 mag!
The new distance modulus ($\sim$900~kpc) would 
be in excellent agreement with D$_{WLM}$ for both stars.   
Such a low mass is unlikely for the atmospheric parameters derived 
for this star though, even if the parameters for this star are 
less certain.    An intermediate solution would be
more satisfactory, e.g., if WLM-31 has a mass near 9 M$_\odot$ and
a slightly higher gravity near \logg=1.8 (or only 1.5\,$\sigma_{\logg}$),
then the distance modulus would be in good agreement ($\sim$900 kpc)
and these values are within reasonable uncertainties in this analysis.
Without being able to quantify the foreground red giant more accurately,
then we will not speculate further on the distance to WLM-31.

We have also looked for the nebular/interstellar features in the WLM spectrum,
such as interstellar NaD lines from WLM, which would be a direct confirmation
of the membership of these stars.  No NaD (nor Ca H/K) lines at the WLM 
\HI\ radial velocity were found, however this is consistent with an
upper-limit to the \NaI\ interstellar abundance.  Assuming a 
limiting W$_\lambda$ of 30 m\AA, and estimating a hydrogen column density 
from its \HI\ 21-cm flux, 300 Jy\,\kms (Huchtmeier \& Richter 1986) 
suggests that N(\NaI)/N(\HI) $\le$ 4x10$^{-9}$.   This upper limit falls 
just above the Galactic relationship determined by Hobbs (1974), 
thus our non-detection of NaD does {\it not} suggest that these are
foreground objects. 
However, the raw 2-D spectral images of WLM-15 {\it do} show 
diffuse H$\alpha$ emission centered around the \HI\ radial velocity 
of WLM ($-$123 $\pm$3 \kms, Huchtmeier \& Richter 1986).   
The diffuse emission is nearly symmetric on either side 
of the stellar spectrum.  The FWHM is 57 \kms, corresponding to 
a (1$\sigma$) velocity dispersion of 24 \kms.    This is in 
excellent agreement with the thermal velocity (22 \kms) of the 
warm interstellar medium (assuming a temperature of 10,000\,K).   
Thus, diffuse H$\alpha$ emission from the WLM dwarf galaxy is 
detected in the spectrum of WLM-15.

Finally, the possibility that the A-supergiants are foreground post-AGB 
stars must be considered.  Post-AGB stars can resemble A-type supergiants
(e.g., Venn \etal 1998), 
typically with low iron-group abundances, higher abundances for 
elements that do not condense into dust grains readily (CNO, S and Zn),
and sharp (shell-like) absorption lines.   The two WLM stars do bear 
some resemblence to post-AGB stars, other than significantly higher 
N and O abundances relative to the iron-group; see Section~\ref{abus}.   
However, it is very 
unlikely that two foreground, Galactic post-AGB stars will be found 
in the WLM field, with the same radial velocity as the H\,I in WLM, 
and with the same abundances as each other.

\subsection{Supergiants in WLM \label{helium}}

The two supergiants presented here are amongst the brightest single stars 
in the galaxy (membership discussed in Section~\ref{membership}).   
That they are low luminosity Ib supergiants begs the 
question ``Where are the Ia supergiants?''.   This same question was 
asked by Sandage \& Carlson (1985) when they could not find Cepheid
variables in WLM with periods of more than 10 days\footnote{Note that
  the missing Cepheid variables would have masses $\ge$7 M$_\odot$, 
  whereas the two supergiants in this paper have M$\sim$12 M$_\odot$}.
The most likely explanation for the missing high mass stars is that
WLM has had an interrupted star formation history during the past
10~Myr (also see the discussion by Skillman \etal 1989a).
The star formation history for WLM has been examined by Dolphin (2000)
from HST WFPC2 imaging (discussed in Section~\ref{sfh}), but the
youngest age bin is 0 to 200~Myr, much larger than the supergiant lifetimes.

Another possibility is that these stars {\it are} the higher mass Ia 
supergiants masquarading as Ib's due to a high surface helium abundance.   
Helium has the effect of increasing the electron density through an
increase in the mean molecular weight in the atmosphere, as noticed
by Kudritzki (1973).     An increase in density primarily mimics an 
increase in the surface gravity, e.g., exhibited in the Balmer line 
profiles through intensified Stark broadening. 
Increasing helium from 9\% to 40\% in the model atmosphere of WLM-15 
mimics a decrease in gravity of only 0.3~dex, i.e., the Balmer lines 
and \MgI/\MgII\ ionization equilibrium  are fully recovered.
This occurs at the same \teff\ (which does imply a shift in the 
\MgI/\MgII\ balance), but surprisingly results in nearly the same LTE
abundances overall; see Table~\ref{unc}. 
The lower gravity would also suggests a higher mass than we have adopted
in Section 3.2;
from the LeJeune \& Schaerer (2001) evolution tracks, the stellar
mass of a Ia supergiant with \logg=1.3 is $\sim$20~M$_\odot$.   
Such a high mass and low gravity would result
in an inconsistent distance modulus for WLM-15 
(D$_{WLM-15} \sim$ 1500~kpc).   We do note however that
some small increase in helium is likely in 
WLM-15 since there is an enrichment in the surface nitrogen abundance, 
see Section~\ref{abus}.   The distance modulus for WLM-31 would
be even larger since the A-supergiant blended in WLM-31 is fainter  
(see the discussion in Section~\ref{membership}).
Thus, we cannot rule out a helium effect, but also we cannot quantify it
at present; fortunately, we notice it would not play any significant 
role in the resulting {\it abundances}.

\section{Abundances \label{abus}}

Elemental abundances have been calculated from an absorption line 
equivalent widths analysis, as well as spectrum synthesis for WLM-15.   
All calculations have been done using a modified version of 
LINFOR\footnote{LINFOR was originally developed by H. Holweger,
  W. Steffen, and W. Steenbock at Kiel University.  Since then,
  it has been upgraded and maintained by M. Lemke, with additional
  modifications by N. Przybilla.   It can be obtained at the 
  following web address: {\it 
  http://www.sternwarte.uni-erlangen.de/pub/MICHAEL/ATMOS-LINFOR-NLTE.TGZ}.}.   
The line list and atomic data were initially adopted from 
previous A-supergiant analyses (Venn 1995a,b, 1999; Venn et al. 2001),
and updated when appropriate, as listed in Table~\ref{lines}.
An attempt has been made to adopt laboratory measurements (e.g.,
O'Brien \etal 1991 for \FeI) and Opacity Project data (e.g., Hibbert 
\etal 1991 for \OI).  Critically examined data were selected next
(e.g., NIST data from Fuhr, Martin, \& Wiese 1988 for \FeII), followed
by the semiempirical values calculated by Kurucz (1988).

Average elemental abundances\footnote{Note that abundances are determined
  from spectral lines of a particular ionization state of an element, but
  the result is the {\it total abundance} of that element since ionization 
  fractions are implicitly calculated in a model atmospheres analysis.} 
for each star are listed in Table~\ref{atms} and shown in Fig.~5.  
Two error estimates are noted per element: the first is the 
line-to-line scatter ($\sigma$) and the second is an {\it estimated}
systematic uncertainty in the atmospheric
parameters (\teff, gravity, and $\xi$).   Abundance uncertainties
due to the model atmospheres are shown in Table~\ref{unc}.
It is clear from this table that \OI, \MgII, \CrII, and \FeII\ 
are amongst the most reliable abundance determinations.  
The systematic error is probably an overestimate since we
have simply added the possible uncertainties in quadrature, not 
accounting for the fact that some \teff-gravity combinations are
excluded by the data.   Scaling the metallicity in the model 
atmosphere's opacity distribution function to 1/3 solar had a 
negligible effect on these abundances ($\le$0.04 dex).
For comparison, solar abundances in Table~\ref{atms} are from 
Grevesse \& Sauval (1998), with the exceptions of 
N and O = 7.80 and 8.66, respectively, from Asplund (2003). 

For WLM-15, spectrum syntheses were performed over several wavelength 
regions, including all of the Mg line regions, as well as the region
around \OI\ $\lambda$6158 and \NI\ $\lambda$7440 (see Figs 6 and 7).  
Macroturbulence equal to 9~\kms\ was adopted based on the instrumental 
resolution and set-up of the UVES spectrograph 
(R=45,000 with a 1.0$''$ slit, then 2x2 binning yields an 
optimal resolution of $\sim$9.4~\kms).   Remaining broadening is 
attributed to $v$sin$i$ ($\sim$7 $\pm$1 \kms).   Spectrum syntheses 
were not performed for WLM-31 because of the continuum offset due to 
the RGB blend (see Section~\ref{recovery}), 
however comparison of WLM-31 and WLM-15 shows 
that the spectra have nearly identically broadening.

{\it Iron-group:}
The iron group abundances are in remarkably good agreement from
lines of \FeI, \FeII, CrI, and CrII, with the mean abundance 
$<$[(Fe,Cr)/H]$>$ = $-$0.36 $\pm$0.17 ({\it $\pm$0.30}) from both stars.
The \FeII\ and \CrII\ abundances (from several absorption lines each) 
are the most reliable, and show only small sensitivities
to the atmospheric parameter uncertainties.   Furthermore, NLTE 
corrections are predicted to be small for these dominant ionization states.  

Sc and Ti can be considered as either iron-group or $\alpha$ elements.
The Ti abundances are determined from several lines and are in excellent 
agreement with iron, [Ti/Fe] = +0.07 $\pm$0.21 ({\it $\pm$0.27}).
The \ScII\ abundances are in fair agreement with the iron-group, 
although from far fewer lines and showing a larger temperature 
sensitivity; furthermore, hyperfine structure 
(the presense of a nuclear magnetic moment for lines of odd
elements of the iron-group) has been neglected.   This should
produce a negligible error in this weak line analysis, and the 
differential \ScII\ results (below) should be more reliable.

{\it The $\alpha$-elements (Mg, O, and Si):}
The mean abundance of magnesium from several lines of two ionic species
in both WLM stars is [Mg/H] = $-$0.62 $\pm$0.09 ({\it $\pm$0.26}),
which is slightly lower than the iron-group result.  \MgII\ is particularly
reliable being rather insensitive to the typical uncertainties in temperature
and gravity, and is found to suffer from negligible NLTE corrections 
(see Section~\ref{atmanal}).
Intriguingly, this contrasts with the oxygen abundance determined from
spectrum synthesis of the $\lambda$6158 feature (see Fig.~6) in WLM-15.
In LTE, the best fit is with 12+log(O/H)=8.60, and we apply a NLTE 
correction of $-$0.15 (Przybilla \etal 2000, based on the atmospheric 
parameters).  Thus, the NLTE abundance listed in Table~\ref{atms} is
[O/H] = $-$0.21 $\pm$0.10 ({\it $\pm$0.05}) (note that 0.10~dex is 
adopted as the line-to-line scatter based on the S/N of the spectrum and 
continuum placement in the spectrum synthesis).
An equivalent width analysis of the 6158~\AA\ line alone results in 
the same abundance.
This result is 2.5 times larger than the magnesium abundance
relative to solar, and
5 times larger than the nebular oxygen abundance!
The oxygen uncertainties in Table~\ref{unc} are determined 
from the W$_\lambda$ analysis of the 6158~\AA\ line abundance.   
Oxygen is not determined in WLM-31;  at 6200~\AA\ the WLM A-supergiant
is expected to contribute only $\sim$13\% to the continuum (from the
Kurucz fluxes, discussed earlier) which is insufficient to measure
the weak \OI\ lines, especially compared to stronger features in that 
region of the red giant spectrum. 

Si appears to be the most underabundant element in these stars
with the mean [Si/H] = $-$0.81 $\pm$0.15 ({\it $\pm$0.28}). 
However, Si abundances have been found to vary by over a factor of 10 
from star to star amongst the A-supergiants in the SMC and the Galaxy 
(Venn 1995b, 1999), including large underabundances not supported by
the other element ratios.  It is not clear if this is due to neglected
NLTE corrections, although it is possible since \SiIII\ would be the 
dominant ionization state at these temperatures.   Thus, the underabundance
of Si in WLM may not be significant.
The Si abundances are discussed further below in Section~\ref{diff}.   

{\it Nitrogen:} 
Nitrogen can be determined in WLM-15 from \NI\ lines near $\lambda$7440
(see Fig.~7).  In LTE, the best fit is for 12+log(N/H) = 8.0, but a NLTE
correction ($-$0.50) is determined here using the departure coefficients 
calculated for the SMC stars AV392 and AV463 by Venn (1999).
These departure coefficients are {\it not} strongly affected
by the new collisional excitation cross-sections 
(calculated by Frost \etal 1998, and discussed for A-type supergiants
by Przybilla 2001b) 
because of the cool temperature of the atmosphere.   Additionally,
while the NLTE corrections are quite large, the differential nitrogen 
abundances with respect to the SMC stars (particularly AV463) should 
be more certain (Section~\ref{diff}).    
The NLTE abundance is [N/H] = $-$0.30 $\pm$0.10 ({\it $\pm$0.04})
(note that 0.10~dex is adopted as the line-to-line scatter based 
on the S/N of the spectrum and continuum placement in the spectrum 
synthesis).
An equivalent width analysis of the 7440 and 7462~\AA\ lines results 
in the same abundance as the spectrum synthesis, thus the    
nitrogen uncertainties in Table~\ref{unc} are determined from these 
two lines.
The nitrogen underabundance is about 1/3 solar, which is much 
higher than the mean nebular nitrogen result from HM95 below 1/10 solar.
However, HM95 note the N/O abundances are quite uncertain, with an upper 
limit of 12+log(N/H) $\le$ 7.30 that is close to our stellar value.

{\it s-process elements:}
One line each of Zr, Ba, and Sr appear to be unblended in the spectra
of both WLM stars.  The \ZrII\ $\lambda$4149 and \BaII\ $\lambda$4934 
results in both stars suggest [s/Fe] ratios near solar; 
the mean value $<$[Zr,Ba/Fe]$>$ = +0.08 $\pm$0.20 ({\it $\pm$0.35}).
This is not surprising if iron is predominantly from SNe Ia and the
s-process abundances from AGB stars, with a similar timescale for
enrichment.   
The \SrII\ $\lambda$4077 line abundance is significantly lower, 
$<$[Sr/Fe]$>$ = $-$1.1, as has been seen in other A-supergiant analyses.   
For example, the mean ratio of [Sr/Fe] in 10 SMC A-supergiants was
also $-$1.0~dex (Venn 1999).  
Belyakova \etal (1999) suggest that \SrII\ abundances in A-stars suffer
from metallicity dependent NLTE effects.   This would also explain
why the mean [Sr/Fe] ratio in Galactic A-supergiants is only $-$0.4~dex
(Venn 1995b).

\section{Differential Abundances \label{diff}}

The absolute abundances determined above are compared to the analyses 
of similar stars in other galaxies in this Section, because systematic
uncertainties in atomic data and model atmosphere assumptions may be
reduced through a differential comparison.   In particular, we will
examine the patterns in abundances and ratios between the two stars in
WLM and previously studied A-supergiants in the Galaxy, SMC, and NGC6822 
(Venn 1995a,b, 1999, Venn \etal 2001).  
Sample spectral regions for the WLM stars and the comparison stars, 
SMC-AV392, SMC-AV463, NGC6822-cc, and Galactic-HD34578, are shown 
in Figs.~6 and 8.   The spectra for NGC6822-cc was taken at the 
Keck telescope with HIRES by J.K. McCarthy (analysed in Venn \etal 2001),
those for the SMC stars were taken at the ESO 3.6-meter telescope
with CASPEC (Venn 1999), and those for the Galactic star 
HD\,34578 were taken with the 2.1-meter telescope at the McDonald 
Observatory with the coud\'e spectrograph (Venn 1995a,b).
All of the spectra have high resolution, 
with 40,000 $\le$ R $\le$ 60,000 per pixel. 
Differential abundances are listed in Tables~\ref{diff-smc} and \ref{diff-oth}
and shown in Fig.~9.  When there are relatively few lines in the analysis 
(e.g., \OI, \MgI, \MgII, \SiII, and the s-process elements), then abundance 
results have been compared line by line and averaged (instead of comparing 
the mean abundances resulting from slightly different line sets).    
 
{\it Comparison with NGC6822-cc:}
The similarities between NGC6822-cc and the two WLM stars in Figs.~6 and 8 
are striking;  the metal lines are extremely sharp\footnote{It is not clear
  why the absorption lines are so sharp in the two WLM stars and NGC6822-cc,
  but not in the Galactic nor SMC comparisons (which have average rotation
  rates of $\sim$20~\kms).   The sharpness implies a very low intrinsic 
  rotation rate since it is unlikely that three of the brightest stars 
  in WLM and NGC6822 also happen to be pole-on rotators.   
  Since the SMC is more metal-poor than these stars, these low rotation
  rates do not appear to be related to metallicity.}
with nearly identical line strengths.
These three stars have nearly the same temperature and metallicity, but 
NGC6822-cc's higher luminosity is apparent from the sharper Balmer line 
in Fig.~8.   Unlike the absolute abundances, these differential abundances
show that the O/Fe and Si/Fe ratios are the same between these two stars. 
Presumably, this is because similar uncertainties in the atmospheric 
analyses cancel.   The differential Mg/Fe ratio is consistent with the
absolute abundance [Mg/Fe] ratio though, and supports that [Mg/Fe] is 
less than the solar ratio in WLM-15.  
(We also note that the differential Sc/Fe 
appears to be severely underabundant, however NGC6822-cc has an unusually 
large Sc abundance which may not be reliable). 

{\it Comparison with SMC-AV463, SMC-AV392, and HD\,34578:}
The differences between the WLM spectra and those for the Galactic 
and SMC comparison stars are also striking in Figs.~6 and 8 (besides the S/N).
It is obvious that the SMC stars are more metal-poor than the 
Galactic star, but due to differences in the broadening parameters
(and, to a lesser degree, slight differences in the atmospheric 
parameters), it is not clear where the WLM stars fall in metallicity 
without the detailed model atmospheres analysis.
Comparing WLM-15 to two SMC A-supergiants (AV392 which is slightly
hotter, and AV463 which is slightly cooler), WLM-15 is clearly more 
metal-rich in iron-group elements by $\sim$0.2~dex.  
However, the O/Fe ratio is the
same between these three stars (and NGC6822-cc as well).
The differential Mg/Fe ratio is suppressed relative to AV463,
and it is marginally lower than in AV392. 

Other differential abundances relative to the SMC stars are also interesting.
The differential Si/Fe ratio supports the absolute abundance result that
Si is suppressed, in contrast to the differential result relative to 
NGC6822-cc.   Since Si abundances vary significantly from star-to-star
in the SMC and Galaxy, this may reflect the true uncertainty in the Si 
abundances (i.e., larger than expected from simple errors in the 
atmospheric parameters, possibly due to neglected NLTE effects).   
It also shows the importance of choosing a comparison star.   
An even larger range in differential abundances is seen for the Sr/Fe
ratios, which is due to the large range in Sr in the SMC stars themselves. 
While this might be 
related to neglected (and metallicity-dependent) NLTE effects, we
note that NLTE corrections are usually similar between similar stars.
Perhaps there is a true range in the strontium abundances in the SMC stars, 
which could be related to inhomogeneous mixing of Sr from AGB stars in the 
SMC's interstellar medium.   Perhaps this is also true for silicon. 

The differential N/Fe ratio is in good agreement between WLM-15 and 
SMC-AV463.  In the SMC, a wide range of nitrogen abundances were found 
and interpreted as evidence for rotational mixing during the main-sequence 
lifetime (which would vary from star to star depending on their 
rotational velocity and mass, Venn 1999a).   
The simple interpretation then would be
that WLM-15 has a mass and had a main-sequence rotation rate 
that are similar to AV463.
 
Finally, a comparison of WLM-15 to the Galactic A-supergiant, HD\,34578,
a slightly less luminous supergiant, shows no significant differences 
from the absolute abundance ratios (relative to solar). 

{\it Summary:} 
The differential abundances suggest that the range in the abundance 
ratios are within the normal range seen in other A-type supergiants 
in other galaxies.   Thus, small differences ($\le$0.2~dex) between 
the abundance ratios are most likely related to model atmosphere analysis 
uncertainties. 
However, the differential comparisons {\do} support that the Mg/Fe 
ratios are less than solar in WLM, and that the O/Fe ratio in WLM-15 
may not be significantly above solar as suggested by the absolute 
abundances.  Thus, the O/Mg ratio does appear higher in WLM-15 than
the comparison stars.
Oxygen in the stars and nebulae in the SMC and NGC6822 were
in excellent agreement (discussed by Venn 1999, Venn \etal 2001).
Thus, the discrepancy in oxygen between the stellar and nebular abundances 
in WLM is significant.   
Finally, there is a large range seen in the differential abundances of 
Si/Fe and Sr/Fe (it is not clear if this reflects a true scatter 
in the star-to-star abundances within a galaxy), while the {\it lack} of
a large difference in the N/Fe ratios between WLM-15 and SMC-AV463 
suggest similar main-sequence rotational mixing histories.

\section{Discussion }

Part of the motivation for determining the stellar abundances
in WLM has been to look for variations in the [O/Fe] ratio between
different dwarf irregular galaxies.    Differences in [O/Fe] imply
significantly different star formation histories (Gilmore \& Wyse 1991)
or differences in infall or outflow (assuming constant IMFs and
stellar yields).   Therefore, variations in the O/Fe ratios between 
galaxies seem likely.

\subsection{The Metallicity and Oxygen Abundances in WLM}

The metallicity determined here for the two young ($\le$10 Myr) supergiants 
in WLM is [Fe/H] = $-$0.38 $\pm$0.20 ({\it $\pm$0.29}).
Reliable magnesium abundances are determined from several lines of two
ionization states in both stars resulting in 
[Mg/H] = $-$0.62 $\pm$0.09 ({\it $\pm$0.26}).
These results suggest that the stars are significantly more metal-rich
than the reported nebular oxygen abundances,
12+log(O/H) = 7.77 $\pm$0.17 from HM95, or [O/H] = $-$0.89.
Oxygen is determined in only one WLM supergiant, WLM-15, from 
spectrum synthesis of the \OI\ 6158 feature, resulting in
[O/H] = $-$0.21 $\pm$0.10 ({\it $\pm$0.05}) (adopting the 
fitting accuracy for the line scatter uncertainty as 
mentioned previously in Section~\ref{abus}), which is even more
discrepant with the nebular oxygen abundance.
Why are the stellar and nebular abundances different 
by 0.68~dex ($>$3$\sigma$)?   
This seems impossible astrophysically since these stars are 
young and presumably formed from the present-day interstellar 
medium sampled by the \HII\ regions.
Possible mechanisms to create an offset are discussed below.

\subsubsection{Could the nebular measurements of O/H be too low?}

Two of the 21 \HII\ regions in WLM (WLM-HM7 and WLM-HM9) are bright 
enough that HM95 were able to detect the [\OIII] $\lambda$4363, but
only WLM-HM7 has a firm measurement, where I($\lambda$4363)/I(H$\beta$) = 
0.09 $\pm$0.01, and thus direct measure of the temperature in the 
O$^{++}$ regions.  The line ratio in WLM-HM9 is I($\lambda$4363)/I(H$\beta$) 
= 0.05 $\pm$0.06.  The only other measurements of abundances in WLM 
are for two \HII\ regions by Skillman \etal (1989a); 
one of these is in common with HM95 (WLM\#1 = WLM-HM9), and again 
the [\OIII] $\lambda$4363 line was detected with a large uncertainty 
(line ratio relative to H$\beta$ was 0.08 $\pm$0.06), and similar
for a second \HII\ region (WLM\#2 = WLM-HM2, with an [\OIII] line ratio
of 0.15 $\pm$0.12). 
Although all four oxygen abundance estimates are in agreement, 
considering the significant uncertainties in the nebular temperatures,
it remains plausible that the nebular oxygen abundances in WLM could
be higher than the existing estimates by as much as 0.2~dex.   Additionally,
low surface brightness \HII\ regions may have relatively large temperature
fluctuations (e.g., Peimbert 1993, Esteban 2002) which would lead to
underestimates of the true nebular oxygen abundances.
New observations of the \HII\
regions in WLM are needed to examine this further. 

   The low value of O/H is consistent with the non-detection of CO
and the metallicity-luminosity relationship for dwarf irregular
galaxies.  Taylor \& Klein (2001) found no CO emission in WLM,
which is consistent with an oxygen abundance of 12+log(O/H) $\le$7.9.   
Other, more distant, galaxies with oxygen abundances above 8.1 
have been detected in CO.    The metallicity-luminosity relationship
for dwarf irregulars is shown in Fig.~10 as the oxygen abundance
versus blue magnitudes M$_B$ from Richer \& McCall (1995).
The stellar oxygen abundance for WLM-15 lies well above this relationship.
The mean [Mg/H] underabundance is in better agreement though (plotting 
12+log(O/H)$_\odot$ + [Mg/H]); in fact it is in excellent agreement with 
NGC6822, which also lies above the relation.   Thus, if the stellar
[Mg/H] ratio is a better estimate of the $\alpha$ element abundances in 
WLM, then its position is consistent with the scatter in the 
metallicity-luminosity relationship.   

\subsubsection{Could the stellar measurements of O/H be too high?}

To reduce the stellar oxygen abundance in WLM-15 to the nebular value 
requires reducing the temperature by $\sim$800~K with no change in 
gravity, or increasing \logg\ by $\sim$1.0 with no change in temperature.   
These would be very large changes indeed; 4$\sigma$ in the temperature
of WLM-15, or 10$\sigma$ in its gravity.  These changes would not reproduce 
the Balmer lines nor \MgI/\MgII\ (and \FeI/\FeII) ionization equilibrium, 
thus it is not possible to reproduce the nebular oxygen abundance in the 
stellar spectrum of WLM-15 (also see Fig.~6). 
To further investigate 
the stellar oxygen abundances in the WLM dwarf galaxy will require 
high resolution observations of additional stars.   In particular, 
analysis of B-type supergiants will permit a more reliable determination 
of the stellar oxygen abundance in WLM from the numerous \OII\ lines.

\subsubsection{Possible mechanisms to create an offset.}

If future observations support the $\Delta$(O/H) = 0.68~dex 
difference between the stellar and nebular oxygen abundances, 
then we consider the following possible scenarios that could 
produce this offset.

{1. {\it Has the nebular oxygen been diluted by infall of metal-poor
         \HI\ gas?}}
Taylor \etal (1995) have shown that \HII\ galaxies can have
companion \HI\ clouds.
The HIPASS survey (Putman \etal 2002) shows there are high velocity 
clouds in the direction of WLM, some with similar radial velocities 
that may suggest an association with WLM.  None are particularly 
close to WLM and all have low \HI\ masses (e.g., HVC~066.1-69.6-172 
is the most significant cloud at 1.25~kpc distance 
with 43\% of the \HI\ flux of WLM), 
but could an \HI\ cloud have recently merged with WLM? 
Gas infall could dilute the interstellar medium in WLM.   If both
the stellar and nebular abundances are accurate, this would have
had to happen in a very short timescale, $\le$10 Myr ($\sim$ age
of the massive supergiants).   Also, the mass of the \HI\ would
have had to be high ($\ge$10$^{6}$ M$_\odot$) to dilute the ISM
by at least 50\%.   These two requirements seem unlikely.   
Unfortunately, no detailed \HI\ mapping is currently available for 
WLM to search for kinematic structures that might be associated 
with a recent merger.  

{2. {\it Large Spatial Variations?}}
The two stars examined here and the two \HII\ regions examined by HM95 
(and an additional \HII\ region by Skillman \etal 1989) 
are located in the south and central regions of the bar of WLM,
see Fig.~1.  Both
stars are on the east side, whereas the nebulae are on the west side.
Could the difference in the abundances be an indication of spatial 
variations in the oxygen abundance in WLM?    The higher oxygen abundances
in the stars on the east side may have formed from a small gas cloud
recently polluted in oxygen over a small volume (e.g., by a local SN II
event which may have also triggered the star formation).   Evidence for
immediate enrichment in oxygen in star forming regions are currently 
inconclusive.  Kobulnicky \& Skillman (1996, 1997) found no evidence for
localized oxygen enrichments from nebular analyses in NGC~1569 nor
NGC~4214 (also, Martin \etal 2002 report higher oxygen abundances in the
hot X-ray gas around NGC~1569 suggesting the newly synthesized material
is injected into the ISM in the hot phase with an uncertain cooling 
and mixing timescale). However, Cunha \& Lambert (1994) find a marginally 
higher oxygen abundance in B-stars in the youngest (Id) subcluster in the 
Orion star forming complex.   
The similarity in the oxygen abundances between the two \HII\ regions
in WLM reported by HM95 (and a third by Skillman \etal 1989) would
suggest that there are no large spatial variations in WLM.   
The similarities in the magnesium and iron-group abundances
in the two A-supergiants in WLM also argues against spatial variations.
Spectral analyses of additional low surface brightness
\HII\ regions and young stars throughout WLM are needed to address this 
problem further.

{3. {\it Changes in the Gas-to-Dust Ratios?}}
Most analyses of gas-phase abundances along sight lines to stars in 
the SMC (Sk~108, Sk~78) have shown broadly similar relative abundance
patterns to those found for the Galactic ISM 
(Welty \etal 1997, Mallouris \etal 2001).
That is, the light elements Mg and Si show mild, monotonic depletions 
with the iron-group elements, and these variations are attributed to 
different environments
in the Galaxy (cold dense cloud, warm diffuse cloud, and halo cloud 
patterns). 
However, recent analyses towards Sk~155 in the wing of the SMC 
shows a significantly different gas-phase abundance pattern
(Welty \etal 2001).   In particular, Mg and Si appear to be relatively
undepleted onto dust grains, whereas the iron-group show moderate
to severe depletions.   This has not been seen in any Galactic sight
line.   Welty \etal  suggest that models of the SMC dust, which currently 
rely heavily on silicates, could be modified to have a dominance of
 oxides and/or metallic grains as an alterative.    Sofia \etal (1994)
also concluded that oxides and/or metallic grains could comprise
a substantial fraction of the Galactic dust.   
If this is extended to WLM, perhaps the low nebular oxygen abundance 
is due to substantial oxygen being locked in dust grains.   However, 
the difference in between the nebular oxygen and stellar oxygen found
here would suggest that $>$50\% of the gas-phase oxygen as been
depleted into dust, a much larger fraction than seen anywhere else
in the Galaxy or SMC.

\subsection{Chemical Evolution of WLM \label{sfh}}

The iron-group abundances in the two WLM stars suggest that this 
galaxy is more chemically evolved than modelled by Dolphin (2000)
who used HST WFPC2 imaging of a small portion of WLM to reconstruct 
its star formation history and chemical evolution from its 
color-magnitude diagram.
Dolphin's analysis suggests that half of WLM's total star formation
(by mass) formed before 9 Gyr ago, then the star formation activity
slowly declined until a recent increase, starting between 1 and 2.5 Gyr 
ago, concentrated in the bar.  He predicts a current metallicity  
of only [Fe/H] = $-$1.08 $\pm$0.18.
Dolphin also notes that this predicted metallicity is in good agreement
with the nebular oxygen value by HM95, thereby suggesting an [O/Fe] 
ratio that is consistent with the solar ratio.
 
It is not clear that our stellar abundances are consistent with a solar
$\alpha$/Fe ratio.  The absolute abundances suggest that [Mg/Fe] is
less than the solar ratio in both stars, whereas [O/Fe] may be slightly 
enhanced in WLM-15.   
The differential abundances agree that Mg/Fe is suppressed,
but that O/Fe is consistent with the stellar ratios in NGC6822 and the 
SMC (where the mean stellar oxygen abundances 
{\it are} in agreement with the nebular results).
Therefore, if we neglect the high stellar oxygen abundance in WLM-15, 
and focus on the more reliable Mg abundance as the $\alpha$ abundance 
indicator, then we find that WLM is the first dwarf irregular galaxy 
that appears to have an $\alpha$/Fe ratio that is less than solar in 
the Local Group.

A less than solar $\alpha$/Fe ratio implies a rapid decline or hiatus in 
the star formation; based on Dolphin's model for WLM, this would have
occurred between 1 and 9 Gyr ago.   During this time, significant SNe~Ia 
enrichment would occur without SNe II enrichment, lowering the $\alpha$/Fe
ratio.  This also implies that little/no mixing of the SNe II products from
the current star formation activity has occurred, but the timescale for 
mixing of SNe II products is not well known and possibly longer than 
$\sim$1~Gyr (e.g., Kobulnicky \& Skillman 1996, 1997; Martin \etal 2002).

Finally, the overall similarities between NGC6822-cc and WLM-15 (supported 
by WLM-31) suggest a very similar chemical evolution history (or at least
chemical evolution end point) for these two galaxies.   
The star formation history in NGC6822 has been examined from its 
ground-based (2.5-meter Isaac Newton Telescope, Canary Islands) 
color-magnitude diagram by Gallart \etal (1996a,b).   Similar to WLM,
Gallart \etal find that NGC6822 formed most of its stars by intermediate
ages, reaching half of its current metallicity $\sim$6~Gyr ago.   They
find a final metallicity of 1/5 solar, which is slightly lower than the
stellar and nebular abundance results in NGC6822 of 1/3 solar.
There was little star formation activity in NGC6822 from $\sim$6~Gyr ago
until recently, $\sim$1~Gyr ago, with particularly strong and localized 
activity in the past 100-200~Myr (leading to spatial abundance variations?).
Dolphin's analysis also shows significant differences in the very 
recent ($\le$1~Gyr) star formation activity between the different WFPC2
WF fields, thus different parts of the WLM galaxy.    
The difference in the Mg/Fe ratios between the stars in these two 
galaxies (see Section~\ref{diff}) may be related to differences
in the length of time of the star formation hiatus in these galaxies
and/or age when SN Ia began to contribute Fe.   The difference in Mg/Fe
could also be related to any number of other chemical evolution parameters
though (e.g., $\alpha$-element losses in a galactic wind).
A study of the
star formation history in WLM from a color-magnitude diagram that covers
the entire galaxy would make it possible to study the progression of 
the recent activity across the galaxy more carefully. 

\section{Conclusions}

We have presented VLT UVES spectroscopy of bright A-type supergiants
in the dwarf irregular galaxy WLM.   Model atmospheres analyses of
the two supergiants, WLM-15 and WLM-31, yield their chemical composition 
(after WLM-31's spectrum was recovered from a blend with a foreground 
red giant).   The abundances are in excellent agreement between the
two stars, yielding a metallicity of [Fe/H] = $-$0.38 $\pm$0.20 
({\it $\pm$0.29}).   This metallicity is supported (within 0.1~dex)
by the abundance results for Na, Ti, and Cr.   
The magnesium abundances are slightly lower,
[Mg/Fe] = $-$0.24 $\pm$0.16 ({\it $\pm$0.28}).
The ratio of the other [$\alpha$/Fe] elements varies from +0.2 (O) 
to $-$0.4~dex (Si), although this pattern changes when differential
abundances (relative to similar stars in NGC6822, the SMC, and
the Galaxy) are examined.   The lower [Mg/Fe] persists, but the 
differential [O/Fe] ratio, and often the [Si/Fe] ratio, is 
consistent with solar.   
If we use the more reliable Mg abundances as 
the $\alpha$ element indicator (more absorption lines analysed, from
two ionization states, with NLTE corrections included, and with
agreement from both stars), then the [$\alpha$(Mg)/Fe] ratio in
WLM suggests this ratio is less than the solar value. 
This is the first time this has been seen in a Local Group galaxy,
and may be related to a hiatus at intermediate ages in its star
formation history.
We also notice that the similarities between WLM-15 and NGC6822-cc
(Venn \etal 2001) are striking, suggesting that WLM and NGC6822 
may have had very similar star formation histories (Dolphin 2000,
Gallart \etal 1996a,b).  The only exception is that the stars in 
WLM have a lower Mg/Fe ratio than those in NGC6822, which may be
related to differences in the length of the star formation hiatus
in these galaxies. 

The metallicity determined from the two WLM stars in this paper 
is {\it not} in agreement with the nebular oxygen abundance from 
\HII\ regions in WLM (Hodge \& Miller 1995, Skillman \etal 1989a).   
The nebular oxygen
abundance is much lower, reported as 12+log(O/H) = 7.77 $\pm$0.17, 
or [O/H] = $-$0.89.   There is no simple explanation for this since
these are young stars that should have formed from the present-day
interstellar medium sampled by the \HII\ regions.
We have considered whether the interstellar medium in WLM could have
been diluted by a companion \HI\ cloud (Taylor \etal 1995), 
but it seems unlikely since a high mass (10$^6$ M$_\odot$) cloud 
would need to have merged and rapidly mixed within the past 10~Myr.   
We have also considered whether gas-phase oxygen could be depleted 
onto oxide dust grains (Welty \etal 2001, Sofia \etal 1994), but over 
half of the interstellar oxygen would have to be locked in dust grains 
which is much more than seen in the Galaxy or SMC.  
Temperature fluctuations (Peimbert 1993, Esteban \etal 2002) would 
move the nebular abundances into better agreement with the stellar results,
particularly if the stars have less than solar $\alpha$/Fe ratios.
The new stellar metallicity presented here pushes WLM above the
metallicity-luminosity relationship (Richer \& McCall 1995)  
observed for traditional dwarf galaxies.
New observations of nebular abundances, stellar abundances, and a
galaxy wide color-magnitude diagram are all modern methods that can
be used to better understand the evolution of WLM.

\acknowledgements
We thank the Paranal Observatory staff for excellent support 
during our visitor run.   Special thanks to Thomas Szeifert
for showing us his FORS2 spectra of WLM supergiants at the telescope,
thanks to Max Pettini for help estimating upper-limits to the nebular 
column densities towards our target stars, and also thanks to Norbert
Przybilla for many stimulating discussions on detailed analyses of
A-type supergiants.    We would also like to acknowledge several
helpful questions and suggestions by the referee (anonymous).
KAV would like to thank the NSF for support through a CAREER award, 
AST-9984073; most of this work was done during a long-term visit at 
the Institute of Astronomy, University of Cambridge, UK.  
ET gratefully acknowledges support from a fellowship 
of the Royal Netherlands Academy of Arts and Sciences, and PATT travel 
support from University of Oxford.    SJS thanks PPARC for Advanced
Fellowship funding.


\clearpage
\begin{deluxetable}{lcccccc} 
\footnotesize
\tablecaption{WLM VLT UVES Observations \label{obs}}
\tablewidth{0pt}
\tablehead{
\colhead{Obj} & \colhead{Date} & \colhead{UT} & \colhead{ $\lambda_c$} &
\colhead{Exptime} & \colhead{Airmass} & \colhead{Seeing} \\[.2ex]
\colhead{}    & \colhead{}  &  \colhead{Begin} & \colhead{ \AA } &
\colhead{sec} & \colhead{} & \colhead{arcsec} 
} 
\startdata
WLM31 & 20Aug2000 & 04:33 & 390+564 & 3600 & 1.19 & 0.89 \\
      &           & 05:34 & 390+564 & 3600 & 1.06 & 0.73 \\
      &           & 06:35 & 390+564 & 3600 & 1.01 & 0.87 \\
      & 21Aug2000 & 02:44 & 390+564 & 3600 & 1.82 & 0.78 \\
WLM15 & 21Aug2000 & 04:02 & 390+564 & 3600 & 1.29 & 1.60 \\
      &           & 06:14 & 390+564 & 2470 & 1.02 & 1.03 \\
      &           & 07:06 & 390+564 & 2470 & 1.02 & 0.68 \\
      &           & 08:09 & 390+840 & 3600 & 1.08 & 0.97 \\
      & 22Aug2000 & 02:52 & 390+840 & 3600 & 1.70 & 0.70 \\
      &           & 03:53 & 390+840 & 3600 & 1.31 & 0.81 \\
\enddata
\tablecomments{Dichroic, 2x2 binning, and 1.0" slit throughout.}
\end{deluxetable}

\clearpage
\begin{deluxetable}{lccccl} 
\footnotesize
\tablecaption{WLM Sample \label{sample}}
\tablewidth{0pt}
\tablehead{
\colhead{Obj} & \colhead{V\tablenotemark{1}} & 
\colhead{(B-V)\tablenotemark{1}} &  \colhead{RV$_{helio}$} & 
\colhead{SpTy} & \colhead{Comment} \\[.2ex] 
\colhead{} & \colhead{} & \colhead{} & \colhead{km/s} 
} 
\startdata
WLM-31 & 18.42 & \phs 0.24    &  $-$117 & A5 Ib & VLT UVES  \\  
WLM-15 & 18.13 & \phs 0.01    &  $-$101 & A5 Ib & VLT UVES  \\
\\
WLM-35 & 18.06 & $-$0.05 & $-$133 & early-B & VLT UVES \\
WLM-10 & 18.20 & $-$0.06 &        & late-B & WHT ISIS\tablenotemark{2} \\
WLM-30 & 18.13 & $-$0.30 &        & early-B    & ESO 3.6m EFOSC\tablenotemark{2} \\
WLM-61 & 17.23 & \phs 0.56    &        & foreground & ESO 3.6m EFOSC\tablenotemark{2} \\
\enddata
\tablenotetext{1}{From Sandage \& Carlson (1985).}
\tablenotetext{2}{Lower resolution spectroscopy by D.J. Lennon and S.J.
Smartt, including ESO 3.6m EFOSC (1989, resolution $\sim$100 \kms) 
and WHT ISIS (July 2000, resolution $\sim$50 \kms).}
\end{deluxetable}

\clearpage
\begin{deluxetable}{lrrlccccc} 
\footnotesize
\tablecaption{Atomic Line Data and LTE Line Abundances \label{lines}}
\tablewidth{0pt}
\tablehead{
\colhead{} & \colhead{} & \colhead{} &
\colhead{} & \colhead{WLM15} & \colhead {WLM15} &
\colhead{WLM31} & \colhead {WLM31} & \colhead{WLM31} \\[.2ex]
\colhead{$\lambda$} & 
\colhead{$\chi$ (eV)} & \colhead{log gf} & 
\colhead{REF}  &
\colhead{EQW} & \colhead{log({\rm X}/H)} &
\colhead{EQW} & \colhead{scaled} & \colhead{log({\rm X}/H)} 
} 
\startdata
N I$_{NLTE}$ \\
\tableline
7442.30 &  10.33 & -0.39 & zhu & 45:+syn &  7.50 & \nodata & \nodata & \nodata \\
7468.31 &  10.34 & -0.19 & zhu & 60:+syn &  7.50  & \nodata & \nodata & \nodata \\
\\
O I \\
\tableline
6155.99 &  10.74 &  -0.67 & op  &  syn  & 8.6  & \nodata & \nodata & \nodata  \\
6156.78 &  10.74 &  -0.45 & op &  syn   & 8.6  & \nodata & \nodata & \nodata  \\
6158.19 &  10.74 &  -0.31 & op & 60+syn & 8.58 & \nodata & \nodata & \nodata  \\
\\
Na I \\
\tableline
5889.95 &   0.00 &   0.11 & wm A & 135+syn & 6.03 & \nodata & \nodata & \nodata \\ 
5895.92 &   0.00 &  -0.19 & wm A &  91+syn & 5.83 & \nodata & \nodata & \nodata \\ 
\\
Mg I$_{NLTE}$ \\
\tableline
3829.36 &   2.71 &  -0.21 & fw B & 125+syn &  6.94  & 135 & 138 & 6.88 \\
4702.99 &   4.35 &  -0.37 & fw C & 35+syn  &  7.02  &  30 &  36 & 6.96 \\
5167.32 &   2.71 &  -0.86 & fw B & 80+syn  &  6.94  &  80 & 110 & 7.18 \\
5172.68 &   2.71 &  -0.38 & fw B & 135+syn &  7.01  &  93 & 127 & 6.86 \\
5183.60 &   2.72 &  -0.16 & fw B & 150+syn &  6.95  & 130 & 178 & \nodata \\
\\
Mg II$_{NLTE}$ \\
\tableline
3848.21 &   8.86 &  -1.56 & fw C & 21+syn  & 6.91   & 25      & 25      &  6.98 \\
4390.57 &  10.00 &  -0.50 & fw D &  38+syn &  6.90  & \nodata & \nodata & \nodata \\
7896.37 &  10.00 &   0.65 & fw C+ & 110+syn &  7.01  & \nodata & \nodata & \nodata \\
\\
Si I \\
\tableline
3905.52 &   1.91 &  -1.09 & fw C  &     99 & 6.95 & 90 & 92 & 6.74 \\
\\
Si II \\
\tableline
3853.66 &   6.86 &  -1.60 & fw E  &  71 & 6.85  &  51 &  52 & 6.51 \\
3856.02 &   6.86 &  -0.65 & fw D+ & 142 & 6.91  & 140 & 143 & 6.64 \\
3862.59 &   6.86 &  -0.90 & fw D+ & 110 & 6.70  & 102 & 104 & 6.45 \\
4130.89 &   9.84 &   0.46 & fw C  &  58 & 6.39  &  85 &  91 & 6.74 \\
5041.02 &  10.07 &   0.17 & fw D+ &  56 & 7.01  &  37 &  48 & 6.80 \\
5055.98 &  10.07 &   0.44 & fw D+ &  45 & 6.54  &  27 &  35 & 6.29 \\
\\
\\
Sc II \\
\tableline
4314.08 &  0.62 &  -0.10 & mfw D & 96 & 2.45 &  80 &  90 & 2.27 \\
4320.73 &  0.61 &  -0.21 & fmw D- & 84 & 2.42 & 104 & 117 & 2.60 \\
4325.00 &  0.60 &  -0.44 & mfw D & 73 & 2.52 & 101 & 113 & 2.79 \\
4374.46 &  0.62 &  -0.42 & mfw D & 83 & 2.62 &  77 &  88 & 2.56 \\
4415.56 &  0.60 &  -0.64 & k88 & 45 & 2.38 &  85 &  98 & 2.85 \\
\\
Ti II \\
\tableline
3932.02 &   1.13 &  -1.78 & mfw D  &     81 & 4.69  &  63 &  64 & 4.42 \\
4012.40 &   0.57 &  -1.61 & mfw C &    112 & 4.49  & 134 & 137 & 4.54 \\
4025.13 &   0.61 &  -1.98 & mfw D- &     82 & 4.51  &  79 &  81 & 4.41 \\
4028.34 &   1.89 &  -1.00 & mfw D &     94 & 4.59  & 120 & 122 & 4.72 \\
4053.81 &   1.89 &  -1.21 & mfw D &     83 & 4.66  &  98 & 101 & 4.72 \\
4161.54 &   1.08 &  -2.36 & mfw D &     35 & 4.62  &  62 &  66 & 4.96 \\
4163.63 &   2.59 &  -0.40 & mfw D &    117 & 4.76  & 120 & 127 & 4.63 \\
4171.92 &   2.60 &  -0.56 & mfw D &    112 & 4.86  & 108 & 114 & 4.67 \\
4287.87 &   1.08 &  -2.02 & mfw D- &     86 & 4.89  &  75 &  84 & 4.77 \\
4301.92 &   1.16 &  -1.16 & mfw D- &    109 & 4.37  & 118 & 133 & 4.40 \\
4312.86 &   1.18 &  -1.16 & mfw D- &    148 & 4.96  & 140 & 158 & 4.67 \\
4316.80 &   2.05 &  -1.42 & mfw D- &     60 & 4.68  &  33 &  37 & 4.35 \\
4320.96 &   1.16 &  -1.87 & fmw D- &     66 & 4.57  & 107 & 120 & 4.99 \\
4330.25 &   2.05 &  -1.52 & mfw D &     30 & 4.36  &  42 &  48 & 4.59 \\
4350.83 &   2.06 &  -1.40 & mfw D &     29 & 4.23  &  30 &  34 & 4.29 \\
4367.66 &   2.59 &  -1.27 & mfw D- &     67 & 4.99  &  64 &  73 & 4.98 \\
4374.82 &   2.06 &  -1.29 & mfw D &     55 & 4.49  & \nodata & \nodata & \nodata \\
4386.84 &   2.60 &  -1.26 & mfw D- &     57 & 4.87  &  35 &  40 & 4.61 \\
4394.05 &   1.22 &  -1.59 & mfw D- &     65 & 4.32  &  68 &  78 & 4.38 \\
4395.85 &   1.24 &  -2.17 & fmw D- &     56 & 4.80  & \nodata & \nodata & \nodata \\
4399.77 &   1.24 &  -1.27 & fmw D- &    115 & 4.60  & 141 & 162 & 4.85 \\
4407.68 &   1.22 &  -2.47 & fmw D- &     21 & 4.53  &  34 &  39 & 4.83 \\
4417.72 &   1.15 &  -1.43 & mfw D- &    104 & 4.55  & 140 & 161 & 4.93 \\
4418.33 &   1.24 &  -2.45 & mfw D- &     32 & 4.75  &  47 &  54 & 5.01 \\
4441.73 &   1.18 &  -2.41 & mfw D- &     44 &	4.84  & \nodata & \nodata & \nodata \\
4450.48 &   1.08 &  -1.45 & mfw D- &     84 & 4.28  & 135 & 157 & 4.85 \\
4464.45 &   1.16 &  -2.08 & mfw D- &     65 & 4.76  &  82 &  94 & 4.96 \\
4501.27 &   1.12 &  -0.75 & mfw D- &    150 & 4.49  & 157 & 181 & \nodata \\
4779.99 &   2.05 &  -1.37 & fmw D- &    70  &	4.71  &  59 &  73 & 4.67 \\
4805.09 &   2.06 &  -1.12 & fmw D- &    95  & 4.75  &  93 & 114 & 4.79 \\
4874.01 &   3.09 &  -0.79 & mfw D &    59  & 4.73  &  60 &  74 & 4.83 \\
4911.19 &   3.12 &  -0.33 & mfw D &    52  & 4.21  &  48 &  60 & 4.25 \\
5013.68 &   1.58 &  -1.94 & k88   &    32  & 4.44  &  20 &  26 & 4.32 \\
5129.15 &   1.89 &  -1.40 & mfw D- & 76  & 4.67  &  39 &  53 & 4.37 \\
5154.07 &   1.57 &  -1.92 & mfw D- & 46  & 4.61  & \nodata & \nodata & \nodata \\
5185.91 &   1.89 &  -1.35 & mfw D  & 42  & 4.22  & \nodata & \nodata & \nodata \\
5188.68 &   1.58 &  -1.22 & mfw D- &    116 & 4.72  &  82 & 112 & 4.51 \\
5226.54 &   1.57 &  -1.29 & mfw D- &    87  & 4.45  & \nodata & \nodata & \nodata \\
5381.02 &   1.57 &  -2.08 & mfw D &    28  & 4.49  & \nodata & \nodata & \nodata \\
5418.75 &   1.58 &  -2.00 & k88  &    27  & 4.39  & \nodata & \nodata & \nodata \\
\\
Cr I \\
\tableline
4254.33 &   0.00 &  -0.11 & mfw B &     44 & 5.20 &  67 &  74 & 5.46 \\
4274.80 &   0.00 &  -0.23 & mfw B &     57 & 5.49 &  60 &  66 & 5.50 \\
5206.04 &   0.94 &   0.02 & mfw B &    30  & 5.45 & \nodata & \nodata & \nodata \\
5208.42 &   0.94 &   0.16 & mfw B &    35  & 5.40 & \nodata & \nodata & \nodata \\
\\
Cr II \\
\tableline
4111.00 &   3.74 &  -1.92 & k88 &     39 & 5.20 &  42 &  44 & 5.24 \\
4145.78 &   5.32 &  -1.16 & k88 &     27 & 5.28 & \nodata & \nodata & \nodata \\
4224.86 &   5.33 &  -1.06 & k88 &     30 & 5.24 & \nodata & \nodata & \nodata \\
4242.36 &   3.87 &  -1.17 & sl &    101 & 5.32 &  84 &  92 & 5.07 \\
4252.63 &   3.86 &  -2.02 & k88 &     30 & 5.23 &  33 &  36 & 5.30 \\
4261.91 &   3.86 &  -1.53 & k88 &     69 & 5.28 &  97 & 107 & 5.56 \\
4275.57 &   3.86 &  -1.71 & k88 &     60 & 5.35 &  53 &  58 & 5.27 \\
4284.19 &   3.85 &  -1.86 & k88 &     39 & 5.21 &  50 &  55 & 5.38 \\
4634.10 &   4.07 &  -1.24 & mfw D &   77 & 5.20 & \nodata & \nodata & \nodata \\
4824.13 &   3.87 &  -1.22 & mfw D &  113 & 5.48 & 110 & 136 & 5.51 \\
4836.22 &   3.86 &  -2.25 & mfw D &   36 & 5.54 & \nodata & \nodata & \nodata \\
4848.24 &   3.86 &  -1.14 & mfw D &   90 & 5.10 &  99 & 123 & 5.30 \\
4876.41 &   3.86 &  -1.46 & mfw D &    85  & 5.36 &  78 &  98 & 5.38 \\
4884.61 &   3.86 &  -2.13 & k88 &    35  & 5.40 & \nodata & \nodata & \nodata \\
5237.33 &   4.06 &  -1.16 & mfw D &    79  & 5.13 & \nodata & \nodata & \nodata \\
5274.96 &   4.05 &  -1.29 & k88 &    57  & 5.00 & \nodata & \nodata & \nodata \\
5334.87 &   4.07 &  -1.56 & k88 &    60  & 5.31 & \nodata & \nodata & \nodata \\
\\
Fe I \\
\tableline
3787.88 &   1.01 &  -0.84 & ob &     60 & 6.80 & \nodata & \nodata & \nodata \\
3820.43 &   0.86 &   0.16 & ob &    158 & 7.22 & \nodata & \nodata & \nodata \\
3825.88 &   0.91 &  -0.03 & ob &    145 & 7.19 & \nodata & \nodata & \nodata \\
3859.91 &   0.00 &  -0.71 & fmw B+ &    144 & 7.18 & 166 & 169 & \nodata \\
3895.66 &   0.11 &  -1.67 & fmw B+ &     71 & 7.11 &  89 &  91 & 7.19 \\
3920.26 &   0.12 &  -1.75 & fmw B+ &     55 & 7.00 &  80 &  82 & 7.19 \\
3922.91 &   0.05 &  -1.65 & fmw B+ &     76 & 7.10 & 105 & 107 & 7.27 \\
3927.92 &   0.11 &  -1.59 & fmw C &     56 & 6.84 &  86 &  88 & 7.08 \\
3930.30 &   0.09 &  -1.59 & fmw C &     91 & 7.25 &  91 &  93 & 7.11 \\
4005.24 &   1.56 &  -0.61 & fmw B+ &     58 & 6.90 &  96 &  98 & 7.20 \\
4021.87 &   2.76 &  -0.66 & fmw C+ &     34 & 7.44 &  43 &  44 & 7.53 \\
4045.81 &   1.48 &   0.28 & ob &    130 & 6.93 & 138 & 142 & \nodata \\
4071.74 &   1.61 &  -0.02 & fmw B+ &     88 & 6.70 & 106 & 111 & 6.76 \\   
4132.06 &   1.61 &  -0.67 & ob &     67 & 7.08 &  98 & 105 & 7.34 \\
4143.87 &   1.56 &  -0.51 & ob &     74 & 6.97 &  76 &  81 & 6.92 \\
4187.80 &   2.43 &  -0.55 & fmw B+ &     45 & 7.25 &  36 &  39 & 7.11 \\
4198.30 &   2.40 &  -0.72 & fmw B+ &     30 &	7.17 &  25 &  27 & 7.06 \\
4199.10 &   3.05 &   0.25 & fmw C &     47 & 6.91 &  53 &  58 & 6.96 \\
4202.03 &   1.48 &  -0.71 & fmw B+ &     53 & 6.85 &  65 &  71 & 6.96 \\
4233.60 &   2.48 &  -0.60 & fmw B+ &     57 & 7.49 &  60 &  66 & 7.50 \\ 
4235.94 &   2.43 &  -0.34 & fmw B+ &     50 & 7.10 &  60 &  66 & 7.20 \\
4250.12 &   2.47 &  -0.40 & fmw B+ &     37 & 7.01 &  60 &  66 & 7.29 \\
4250.79 &   1.56 &  -0.72 & fmw D- &     45 & 6.81 &  84 &  93 & 7.23 \\
4260.47 &   2.40 &  -0.02 & fmw D &     98 & 7.35 &  78 &  87 & 7.06 \\
4271.15 &   2.45 &  -0.35 & fmw B+ &     45 & 7.06 &  69 &  77 & 7.33 \\
4271.76 &   1.48 &  -0.16 & fwm B+ &    101 & 6.88 & 158 & 175 & \nodata \\
4282.40 &   2.18 &  -0.82 & fmw C+ &     35 & 7.19 &  39 &  43 & 7.25 \\
4325.76 &   1.61 &   0.01 & ob &    120 & 7.06 & 109 & 123 & 6.80 \\
4383.54 &   1.48 &   0.21 & ob &    110 & 6.62 & 174 & 198 & \nodata \\
4415.12 &   1.61 &  -0.61 & fmw B+ &     83 & 7.18 &  77 &  89 & 7.10 \\
4459.12 &   2.18 &  -1.28 & fmw B+ &     23 & 7.41 & \nodata & \nodata & \nodata \\
4466.55 &   2.83 &  -0.60 & fmw C+ &     26 & 7.24 & \nodata & \nodata & \nodata \\
4476.02 &   2.85 &  -0.73 & k88 &     32 & 7.50 & \nodata & \nodata & \nodata \\
4494.56 &   2.20 &  -1.14 & fmw B+ &     30 & 7.42 &  25 &  29 & 7.36 \\
4871.32 &   2.86 &  -0.42 & k88 &     35 & 7.23 &  58 &  72 & 7.60 \\
4872.14 &   2.88 &  -0.62 & k88 &     19 & 7.12 &  30 &  37 & 7.42 \\
4890.75 &   2.88 &  -0.42 & fmw C+ &     30 & 7.15 & \nodata & \nodata & \nodata \\
4891.50 &   2.85 &  -0.11 & ob &     50 & 7.12 &  50 &  63 & 7.19 \\
4919.00 &   2.87 &  -0.37 & fmw C+ & 37 & 7.21 &  48 &  60 & 7.44 \\
4920.50 &   2.83 &   0.06 & fmw C+ & 70 & 7.18 &  80 & 100 & 7.36 \\
4957.60 &   2.84 &  -0.49 & fmw D &  50 & 7.49 & \nodata & \nodata & \nodata \\ 
5192.34 &   3.00 &  -0.52 & k88 &    23 & 7.19 & \nodata & \nodata & \nodata \\
5364.86 &   4.45 &   0.22 & fmw D &  22 & 7.42 & \nodata & \nodata & \nodata \\
5371.49 &   0.96 &  -1.64 & wm B+ &  22 & 6.84 & \nodata & \nodata & \nodata \\
5383.37 &   4.31 &   0.50 & fmw C+ & 50 & 7.51 & \nodata & \nodata & \nodata \\
5405.77 &   0.99 &  -1.84 & wm B+ & 28 & 7.19 & \nodata & \nodata & \nodata \\
5429.70 &   0.96 &  -1.88 & wm B+ & 25 & 7.15 & \nodata & \nodata & \nodata \\
\\
Fe II \\
\tableline
3783.35 &   2.27 &  -3.16 & k88 &     95 & 6.87 & 120 & 122 & 6.97 \\
3945.21 &   1.70 &  -4.25 & k88 &     55 & 7.03 &  58 &  59 & 7.02 \\
4057.46 &   7.27 &  -1.55 & k88 &     27 & 7.42 & \nodata & \nodata & \nodata \\
4128.75 &   2.58 &  -3.76 & fmw D &     70 & 7.31 &  57 &  60 & 7.12 \\
4258.15 &   2.70 &  -3.40 & fmw D &     68 & 7.02 & 107 & 117 & 7.38 \\
4273.32 &   2.70 &  -3.34 & fmw D &     75 & 7.05 &  81 &  90 & 7.07 \\
4296.57 &   2.70 &  -3.10 & mfw D &     97 & 7.05 & 115 & 128 & 7.19 \\
4303.17 &   2.70 &  -2.49 & fmw D &    149 & 7.22 & \nodata & \nodata & \nodata \\
4354.34 &   7.65 &  -1.40 & k88 &     22 & 7.41 & \nodata & \nodata & \nodata \\
4369.41 &   2.78 &  -3.66 & fmw D &     45 & 7.01 &  83 &  95 & 7.49 \\
4385.38 &   2.78 &  -2.57 & fmw D &    140 & 7.19 & \nodata & \nodata & \nodata \\
4472.92 &   2.84 &  -3.43 & fmw D &     75 & 7.19 &  50 &  59 & 6.94 \\
4489.18 &   2.83 &  -2.97 & fmw D &     75 & 6.72 & 114 & 133 & 7.18 \\ 
4491.40 &   2.86 &  -2.69 & fmw C &    131 & 7.20 & 127 & 148 & 7.07 \\
4629.34 &   2.81 &  -2.38 & fw D &   140 & 6.98 & \nodata & \nodata & \nodata \\
4635.32 &   5.96 &  -1.65 & fmw D- &     40 & 7.00 & \nodata & \nodata & \nodata \\
4663.71 &   2.89 &  -4.27 & k88 &     30 & 7.45 & \nodata & \nodata & \nodata \\ 
4666.75 &   2.83 &  -3.33 & fmw D &     74 & 7.06 &  64 &  78 & 7.02 \\
4731.45 &   2.89 &  -3.37 & fmw D &     67 & 7.06 &  74 &  91 & 7.22 \\
4993.36 &   2.81 &  -3.65 & fmw E &     47 & 7.03 & \nodata & \nodata & \nodata \\
5197.57 &   3.23 &  -2.10 & fmw C &    135 & 6.88 & 122 & 167 & 6.91 \\
5234.62 &   3.22 &  -2.05 & fmw C &    129 & 6.74 & \nodata & \nodata & \nodata \\ 
5264.81 &   3.23 &  -3.19 & fmw D &     59 & 7.01 & \nodata & \nodata & \nodata \\
5284.10 &   2.89 &  -3.19 & fmw D &     67 & 6.87 & \nodata & \nodata & \nodata \\
5337.73 &   3.23 &  -3.31 & k88 &     35 & 6.80 & \nodata & \nodata & \nodata \\
5362.86 &   3.20 &  -2.74 & k88 &    105 & 7.09 & \nodata & \nodata & \nodata \\
5425.25 &   3.20 &  -3.36 & fmw D &     53 & 7.09 & \nodata & \nodata & \nodata \\
5534.85 &   3.24 &  -2.92 & fmw D &     85 & 7.06 & \nodata & \nodata & \nodata \\
6147.74 &   3.89 &  -2.46 & fmwy D &     48 & 6.62 & \nodata & \nodata & \nodata \\ 
6149.24 &   3.89 &  -2.77 & fmw D &     55 & 7.02 & \nodata & \nodata & \nodata \\
6238.38 &   3.89 &  -2.48 & k88 &     87 & 7.13 & \nodata & \nodata & \nodata \\
6247.56 &   3.89 &  -2.36 & fmw D &     68 & 6.78 & \nodata & \nodata & \nodata \\
6416.91 &   3.89 &  -2.70 & fmw D &     70 & 7.15 & \nodata & \nodata & \nodata \\
6456.38 &   3.90 &  -2.16 & fmw D &    120 & 7.24 & \nodata & \nodata & \nodata \\
7462.41 &   3.89 &  -2.73 & k88 &     90 & 7.48 & \nodata & \nodata & \nodata \\ 
\\
Sr II \\
\tableline
4077.71 &  0.00  &   0.17 & k88 &    62  &  1.19 & 116 & 121 & 1.67 \\
\\
Zr II \\
\tableline
4149.22 &  0.80  &  -0.03 & k88 &    54  &  2.30 &  70 &  75 & 2.46 \\
\\
Ba II \\
\tableline
4934.08 &  0.00  &  -0.15 & k88 &   30:  &  1.66 &  50 &  63 & 2.03 \\
\enddata
\tablerefs{
Transition Probability References:
fw = Fuhr \& Wiese (1998);
fmw = Fuhr, Martin, \& Wiese (1988);
fmwy = Fuhr \etal (1981);
k88 = Kurucz (1988);
mfw = Martin, Fuhr, \& Wiese (1988);
ob = O'Brian \etal (1991);
op = Opacity Project (Hibbert \etal 1991);
sl = Sigut \& Landstreet (1990);
wm = Wiese \& Martin (1980);
zhu = Zhu \etal (1989).
}
\end{deluxetable}

\clearpage
\begin{deluxetable}{lccc} 
\footnotesize
\tablecaption{Atmospheric Analysis \label{atms}}
\tablewidth{0pt}
\tablehead{
\colhead{} & \colhead {Solar} & \colhead{WLM-15} & \colhead{WLM-31} }
\startdata
\teff &    & 8300 $\pm$200 & 8300 $\pm$300 \\
\logg &    &  1.6 $\pm$0.1 & 1.65 $\pm$0.1  \\
$\xi$   &    &  3 $\pm$1   & 4 $\pm$1      \\
log(L/L$_\odot$)\tablenotemark{a} &  &  4.55 $\pm$0.20 & \nodata \\
R/R$_\odot$\tablenotemark{a}     &   &  91 $\pm$20 & \nodata  \\
Sp. Ty. &                            & A5 Ib &  A5 Ib \\
\\
N I NLTE & 7.80 & 7.50  (syn) {\it $\pm$0.04}         & \nodata  \\
O I NLTE & 8.66 & 8.45  (syn) {\it $\pm$0.05}         & \nodata  \\
Na I     & 6.32 & 5.93 $\pm$0.14 (2) {\it $\pm$0.30}  & \nodata  \\
\\
Mg II NLTE & 7.58 & 6.94 $\pm$0.06 (3) {\it $\pm$0.08} & 6.97 (1) {\it $\pm$0.03} \\
Mg I  NLTE & 7.58 & 6.97 $\pm$0.04 (5) {\it $\pm$0.26} & 6.96 $\pm$0.15 (4) {\it $\pm$0.43} \\
Si II  & 7.56 & 6.73 $\pm$0.23 (6) {\it $\pm$0.14}  &  6.57 $\pm$0.19 (6) {\it $\pm$0.15}   \\
Si I   & 7.56 & 6.95 (1) {\it $\pm$0.31}            &  6.74 (1) {\it $\pm$0.42}   \\
Ti II  & 4.94 & 4.60 $\pm$0.21 (40) {\it $\pm$0.20} &  4.65 $\pm$0.24 (31) {\it $\pm$0.26}  \\
\\
Sc II  & 3.10 & 2.48 $\pm$0.09 (5) {\it $\pm$0.22}  &  2.62 $\pm$0.23 (5) {\it $\pm$0.34}  \\
Cr II  & 5.69 & 5.27 $\pm$0.14 (17) {\it $\pm$0.09} &  5.33 $\pm$0.15 (9) {\it $\pm$0.15}  \\
Cr I   & 5.69 & 5.38 $\pm$0.13 (4)  {\it $\pm$0.29} &  5.48 $\pm$0.03 (2) {\it $\pm$0.45}  \\
Fe II  & 7.50 & 7.06 $\pm$0.21 (35) {\it $\pm$0.16} &  7.12 $\pm$0.17 (13) {\it $\pm$0.19}  \\
Fe I   & 7.50 & 7.13 $\pm$0.22 (47) {\it $\pm$0.30} &  7.18 $\pm$0.22 (30) {\it $\pm$0.44}  \\
\\
Sr II  & 2.92 & 1.21 (1) {\it $\pm$0.34}  & 1.67 (1) {\it $\pm$0.50}  \\
Zr II  & 2.61 & 2.30 (1) {\it $\pm$0.17}  & 2.46 (1) {\it $\pm$0.32}  \\
Ba II  & 2.22 & 1.66 (1) {\it $\pm$0.35}  & 2.03 (1) {\it $\pm$0.50}  \\
\enddata
\tablenotetext{a}{The WLM-15 luminosity and radius are determined from 
the stellar atmospheric parameters and adopting M = 12 M$_\odot$ from 
standard stellar evolution tracks, e.g., Lejeune \& Schaerer (2001).
We suggest that WLM-31 may have a slightly lower mass 
(see Section~\ref{membership}}.
\tablecomments{
Two uncertainties are shown for each elemental abundance, the line-to-line
scatter (also shown are the number of lines used in the abundance), and
an estimate of the systematic error in italics.   The systematic errors
are based on the uncertainties in the atmospheric parameters, tabulated
in Table~\ref{unc}. 
Solar abundances are Grevesse \& Sauval (1998), with the exception 
of N and O from Asplund (2003).
Although elemental abundances are calculated for a particular ionization
state, the results do represent the total abundance of the element (thus,
ionization balance, when available, can be a useful test of the reliability 
of the results). 
} 
\end{deluxetable}

\clearpage
\begin{deluxetable}{lccccc} 
\footnotesize
\tablecaption{Standard LTE Abundance Uncertainties \label{unc}}
\tablewidth{0pt}
\tablehead{
\colhead{} & \colhead{WLM-15} & \colhead{WLM-15\tablenotemark{1}} &
\colhead{WLM-15\tablenotemark{1}} & \colhead{WLM-15\tablenotemark{2}} & 
\colhead{WLM-31}  \\[.2ex]
\colhead{} & \colhead{$\Delta$\teff} & \colhead{$\Delta$\logg} &
\colhead{$\Delta\xi$} & \colhead{Helium} & \colhead{$\Delta$\teff} \\[.2ex] 
\colhead{} & \colhead{=+200\,K} & \colhead{=+0.1} & \colhead{=+1 km/s} &
\colhead{Test} & \colhead{=+300\,K} } 
\startdata
$\Delta$log(NI/H)   & +0.03 & $-$0.02   & $-$0.02 & +0.05 & \nodata \\
$\Delta$log(OI/H)   & +0.02 & \phs 0.00 & $-$0.05 & +0.04 & \nodata \\
$\Delta$log(NaI/H)  & +0.25 & $-$0.06   & $-$0.14 & +0.03 & \nodata \\
$\Delta$log(MgII/H) &  \phs 0.00 & \phs 0.00 & $-$0.08 & +0.03 & +0.03 \\
$\Delta$log(MgI/H)  & +0.20 & $-$0.06   & $-$0.16 & +0.01 & +0.39 \\
$\Delta$log(SiII/H) & $-$0.02 & +0.01   & $-$0.14 & +0.04 & $-$0.05 \\
$\Delta$log(SiI/H)  & +0.28 & $-$0.06   & $-$0.13 & $-$0.02 & +0.40 \\
$\Delta$log(TiII/H) & +0.17 & \phs 0.00 & $-$0.10 & $-$0.06 & +0.24 \\
$\Delta$log(ScII/H) & +0.21 & $-$0.01   & $-$0.08 & $-$0.07 & +0.33 \\
$\Delta$log(CrII/H) & +0.06 & \phs 0.00 & $-$0.07 & $-$0.04 & +0.14 \\
$\Delta$log(CrI/H)  & +0.28 & $-$0.06   & $-$0.03 & $-$0.02 & +0.45 \\
$\Delta$log(FeII/H) & +0.10 &   +0.01   & $-$0.12 & $-$0.03 & +0.15 \\
$\Delta$log(FeI/H)  & +0.28 & $-$0.06   & $-$0.10 & $-$0.02 & +0.42 \\
$\Delta$log(SrII/H) & +0.33 & $-$0.04   & $-$0.07 & $-$0.06 & +0.49 \\
$\Delta$log(ZrII/H) & +0.16 & $-$0.01   & $-$0.05 & $-$0.07 & +0.32 \\
$\Delta$log(BaII/H) & +0.34 & $-$0.06   & $-$0.02 & $-$0.04 & +0.50 \\
\enddata
\tablenotetext{1}{Uncertainties in gravity and microturbulence
for WLM31 are similar to those listed here for WLM15.}
\tablenotetext{2}{An ATLAS9 model atmosphere for
WLM15 with 40\% helium content has been adopted
for this test.  Increasing helium also required
a change in gravity of $\Delta$\logg = $-$0.3 to 
fit the Balmer lines.  No other changes were made
(i.e., to \teff\ or $\xi$).}
\end{deluxetable}

\clearpage
\begin{deluxetable}{lclcl} 
\footnotesize
\tablecaption{Differential Abundance Ratios (SMC) \label{diff-smc}}
\tablewidth{0pt}
\tablehead{
\colhead{} & \colhead{} & \colhead{WLM-15 $-$} & 
                \colhead{} & \colhead{WLM-15 $-$} \\[.2ex]
\colhead{} & \colhead{AV\,392} & \colhead{AV\,392} &
             \colhead{AV\,463} & \colhead{AV\,463} } 
\startdata
Teff       &       8500         & & 8000      \\ 
log g      &        1.7         & &  1.3      \\
$\xi$      &          3         & &  4        \\
Sp.Ty.     &       A3 Ib        & &  A7 Ib    \\
$[$FeII/H$]$ & $-$0.69 (19) & +0.25  & $-$0.60 (33) & +0.16 \\
$[$OI/H$]$   & $-$0.46 (1)  & +0.25  & $-$0.46 (2)  & +0.25 \\ 
\\
$[$NI$_{NLTE}$/FeII$]$ & \nodata & \nodata  & \phs 0.00 (1)\tablenotemark{*} & +0.14 \\ 
$[$OI$_{NLTE}$/FeII$]$ & +0.23 (1) & \phs 0.00 & +0.14 (2) & +0.09 \\ 
$[$MgI$_{NLTE}$/FeII$]$  & $-$0.06 (4) & $-$0.11 & +0.01 (3) & $-$0.18 \\ 
$[$MgII$_{NLTE}$/FeII$]$ & $-$0.06 (1)\tablenotemark{*} & $-$0.14 & +0.10 (1)\tablenotemark{*} & $-$0.30 \\ 
$[$SiII/FeII$]$ & +0.24 (2)\tablenotemark{*} & $-$0.58 & $-$0.10 (2)\tablenotemark{*} & $-$0.29 \\ 
$[$NaI/FeII$]$ & \nodata\tablenotemark{a} & \nodata & $-$0.15 (2)  & +0.20 \\
$[$TiII/FeII$]$ & +0.23 (22)  & $-$0.13 &   +0.14 (32) & $-$0.04 \\
$[$ScII/FeII$]$ & +0.01 (1)\tablenotemark{*}   & $-$0.19 & \nodata\tablenotemark{a} & \nodata \\
$[$CrII/FeII$]$ & +0.09 (12)  & $-$0.07 & $-$0.09 (15) &   +0.11 \\
$[$SrII/FeII$]$ & $-$0.83 (1)\tablenotemark{*} & $-$0.66 & $-$1.11 (1)\tablenotemark{*} & $-$0.16 \\ 
$[$ZrII/FeII$]$ & \nodata & \nodata & $-$0.17 (1)\tablenotemark{*} &   +0.30 \\
$[$BaII/FeII$]$ & \nodata & \nodata & $-$0.06 (1)\tablenotemark{*} & $-$0.06 \\
\enddata
\tablenotetext{*}{When only a few lines are used in the abundance
($\le$3) then only the lines in common in both analyses are compared.}
\tablenotetext{a}{No lines in common.}
\end{deluxetable}

\clearpage
\begin{deluxetable}{lclcl} 
\footnotesize
\tablecaption{Differential Abundance Ratios 
(NGC6822 \& Galactic) \label{diff-oth}}
\tablewidth{0pt}
\tablehead{
\colhead{} & \colhead{} & \colhead{WLM-15 $-$} & 
                \colhead{} & \colhead{WLM-15 $-$} \\[.2ex]
\colhead{} & \colhead{NGC6822-cc} & \colhead{NGC6822-cc} &
             \colhead{HD34578}           & \colhead{HD34578} } 
\startdata
Teff      &       8500         & & 8300      \\ 
log g      &        1.1         & &  1.85     \\
$\xi$      &          6         & &  4        \\
Sp.Ty.     &       A3 Ia        & &  A5 II    \\
$[$FeII/H$]$    & $-$0.40 (20) & $-$0.04  & +0.04 (17) & $-$0.48 \\
$[$OI/H$]$      & $-$0.22 (2)\tablenotemark{*} & +0.01  & $-$0.08 (3)\tablenotemark{a} & $-$0.13 \\
\\
$[$NI/FeII$]$    & \nodata & \nodata & +0.06 (2)\tablenotemark{*} & +0.08 \\
$[$OI$_{NLTE}$/FeII$]$   & +0.18 (2)\tablenotemark{*} &   +0.05  &
   $-$0.12 (3)\tablenotemark{a} &  +0.35 \\ 
$[$MgI$_{NLTE}$/FeII$]$  & +0.12 (2)\tablenotemark{*} & $-$0.28  &
   $-$0.09 (1)\tablenotemark{*} & $-$0.03 \\
$[$MgII$_{NLTE}$/FeII$]$ & +0.16 (1)\tablenotemark{*} & $-$0.40  &
   \nodata\tablenotemark{b} &  \nodata \\
$[$SiII/FeII$]$          & +0.21 (1)\tablenotemark{*} & $-$0.11  &
   \nodata\tablenotemark{b} &  \nodata \\
$[$TiII/FeII$]$ &    +0.01 (13)      &   +0.09 &
   +0.26 (17) & $-$0.16   \\
$[$ScII/FeII$]$ & +0.29 (2)\tablenotemark{*}  & $-$0.60 &
   \nodata\tablenotemark{b} & \nodata  \\
$[$CrII/FeII$]$ & $-$0.07 (10) &   +0.09  & +0.08 (4)  & $-$0.06 \\
\enddata
\tablenotetext{*}{When only a few lines are used in the abundance
($\le$3) then only the lines in common in both analyses are compared.}
\tablenotetext{a}{NLTE correction of $-$0.15 dex has been applied
to the LTE oxygen abundance for HD34578 by Venn 1995b.}
\tablenotetext{b}{No lines in common.}
\end{deluxetable}


\clearpage
\epsscale{0.8}
\plotone{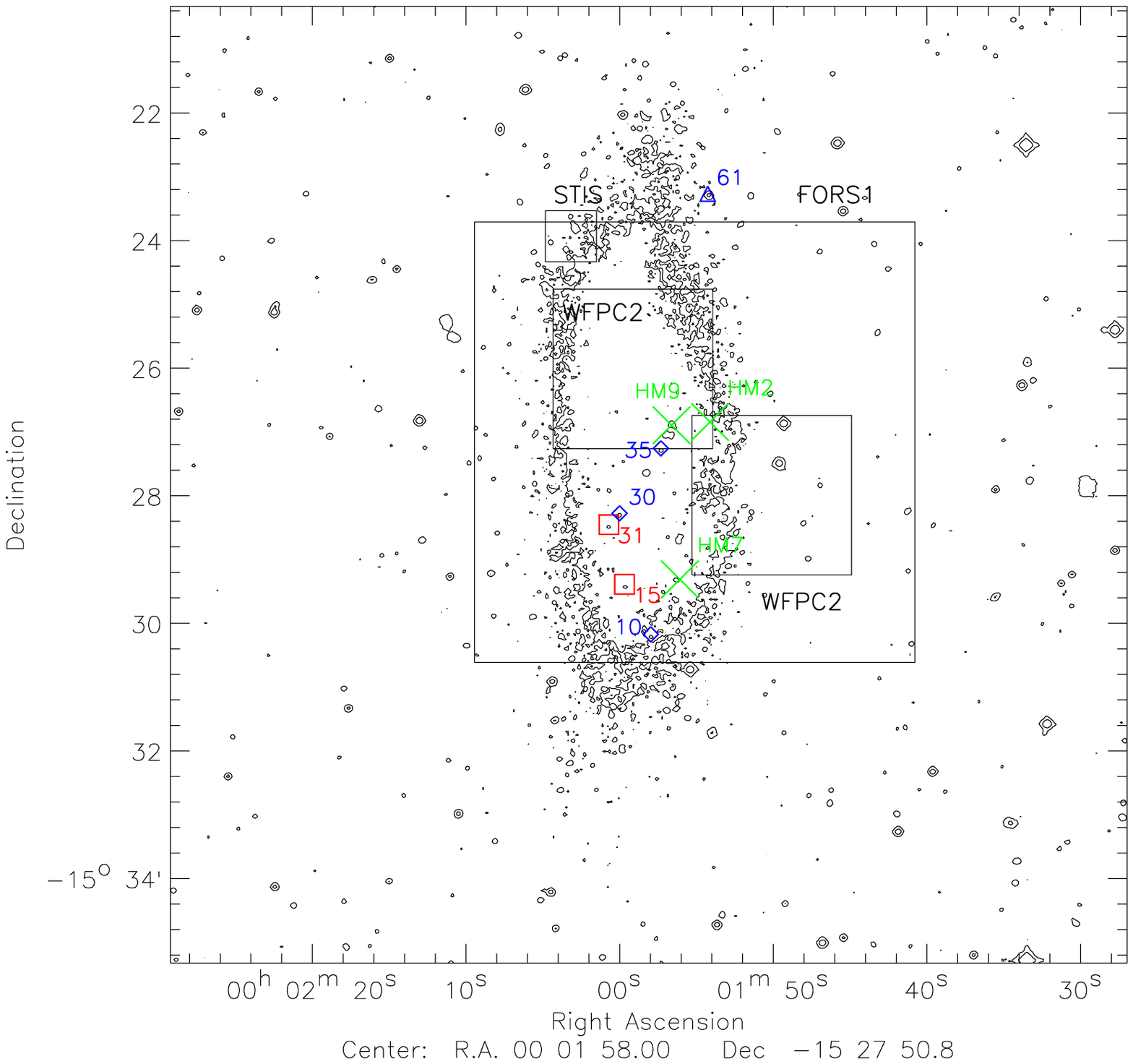}
\figcaption{
The WLM dwarf irregular galaxy.  The two A-supergiants analysed in
this paper are noted (in red), as well as other blue supergiants from 
the original target list (in blue) and listed in Table~\ref{sample}.   
The \HII\ regions where oxygen abundances have been determined from
the [\OIII] $\lambda$4363 line are marked (in green), and the fields 
of view for VLT-FORS1, HST-STIS, and HST-WFPC2 imaging are outlined.  
}
\label{fig1}

\clearpage
\epsscale{0.8}
\plotone{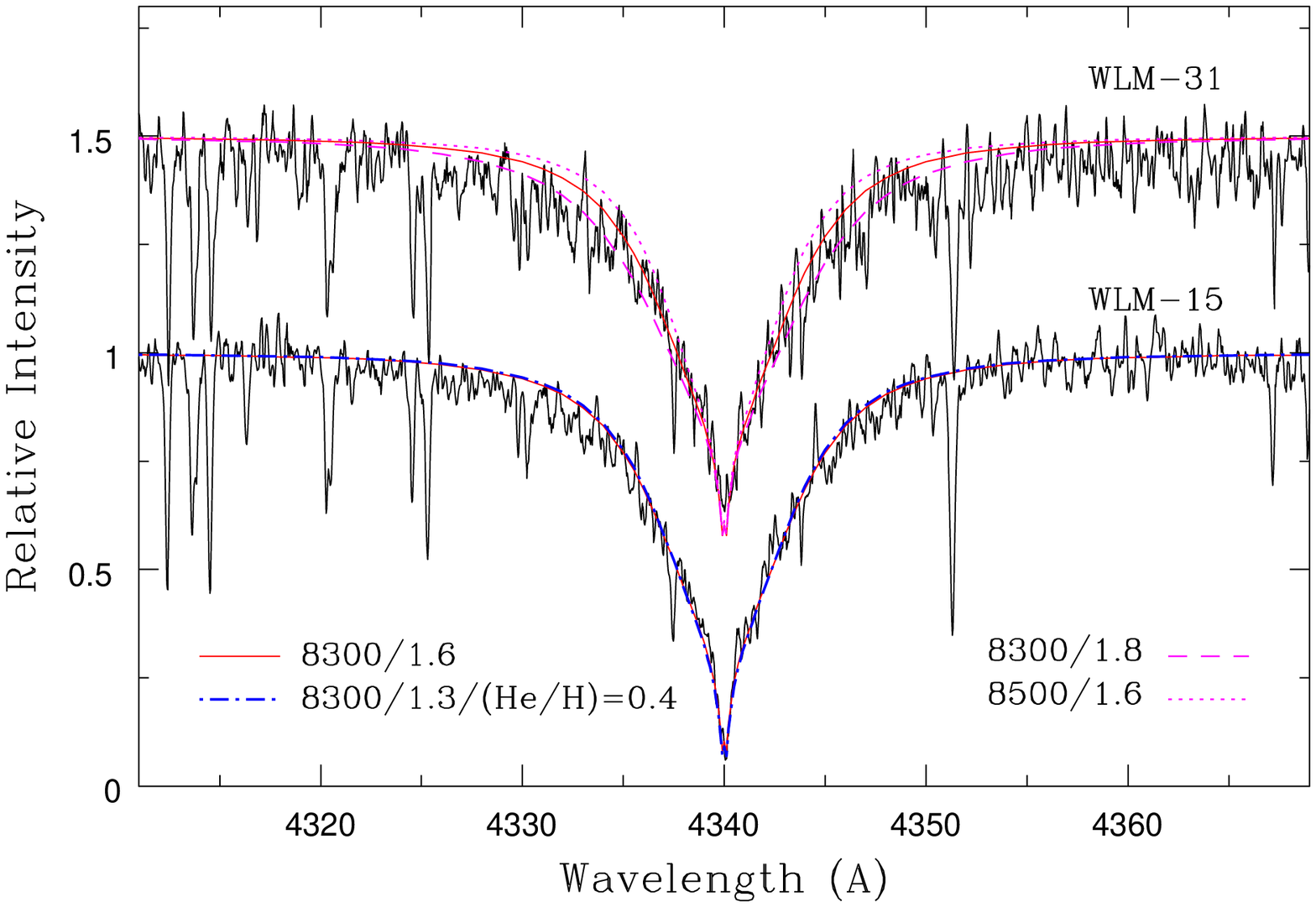}
\figcaption{
H$\gamma$ profile for WLM-15 and WLM-31 and fits from different model
atmospheres.   The model with increased helium abundance and lower
gravity is identical to the best fit model.
}
\label{fig2}

\clearpage
\epsscale{0.8}
\plotone{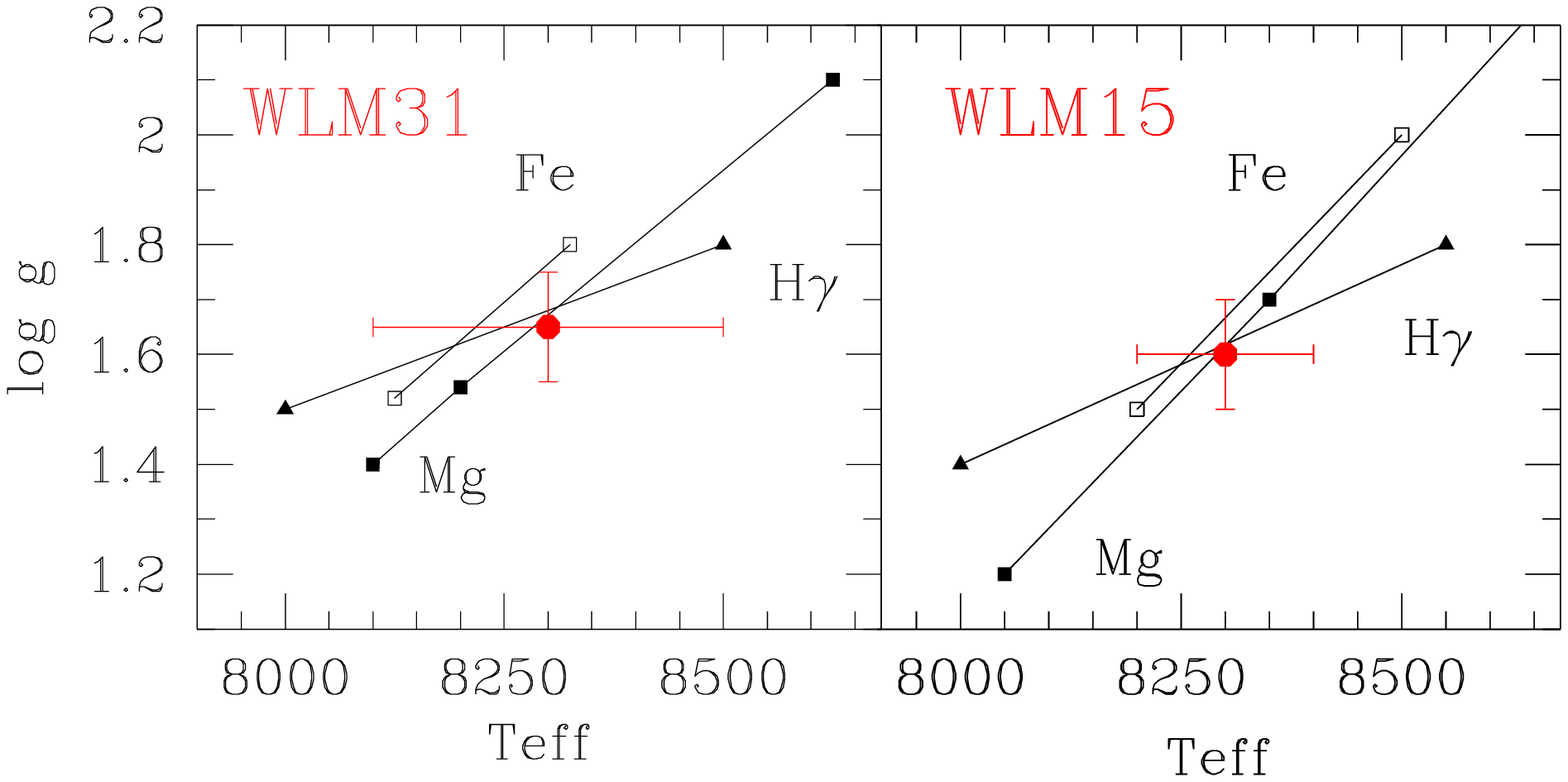}
\figcaption{
Atmospheric parameter selections for WLM-31 and WLM-15 
(solid circles with errorbars). 
H$\gamma$ fits (solid triangles), \MgI/\MgII\ (solid squares),
and \FeI/\FeII (hollow squares) ionization equilibrium are used 
for both stars.   
}
\label{fig3}

\clearpage
\epsscale{0.8}
\plotone{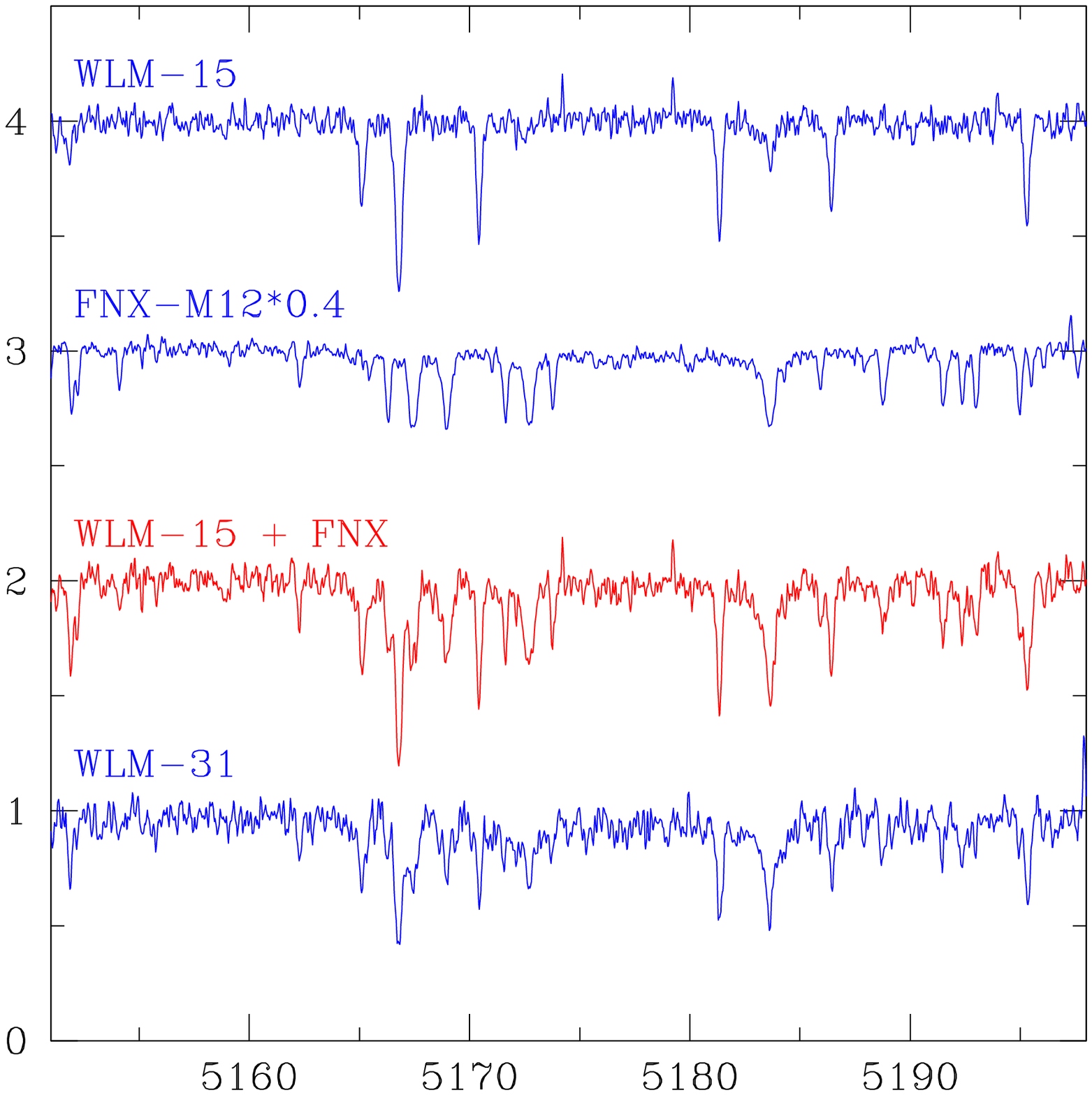}
\figcaption{
Spectrum of WLM-31 around the Mgb lines.  
The spectrum of WLM-15 is combined with that of a metal-poor 
red giant in Fornax (M12, Shetrone \etal 2003) scaled by 40\%, 
to show that the combination is a very good match for the WLM-31 
spectrum.   Note that the Fornax red giant spectrum is only
for illustration; in the text we discuss that the continuum scaling
is independent of gravity near 4250~K, thus the contaminating
star does not need to be a high luminosity red giant.
The Fornax red giant has been corrected to have 
zero radial velocity, whereas the spectra of WLM-31 and
WLM-15 have not been corrected for this illustration.    The 
contribution of the red giant to the WLM-31 spectrum drop off at 
bluer wavelengths becoming negligible near 4000~\AA\ (see text). 
}
\label{fig4}

\clearpage
\epsscale{0.8}
\plotone{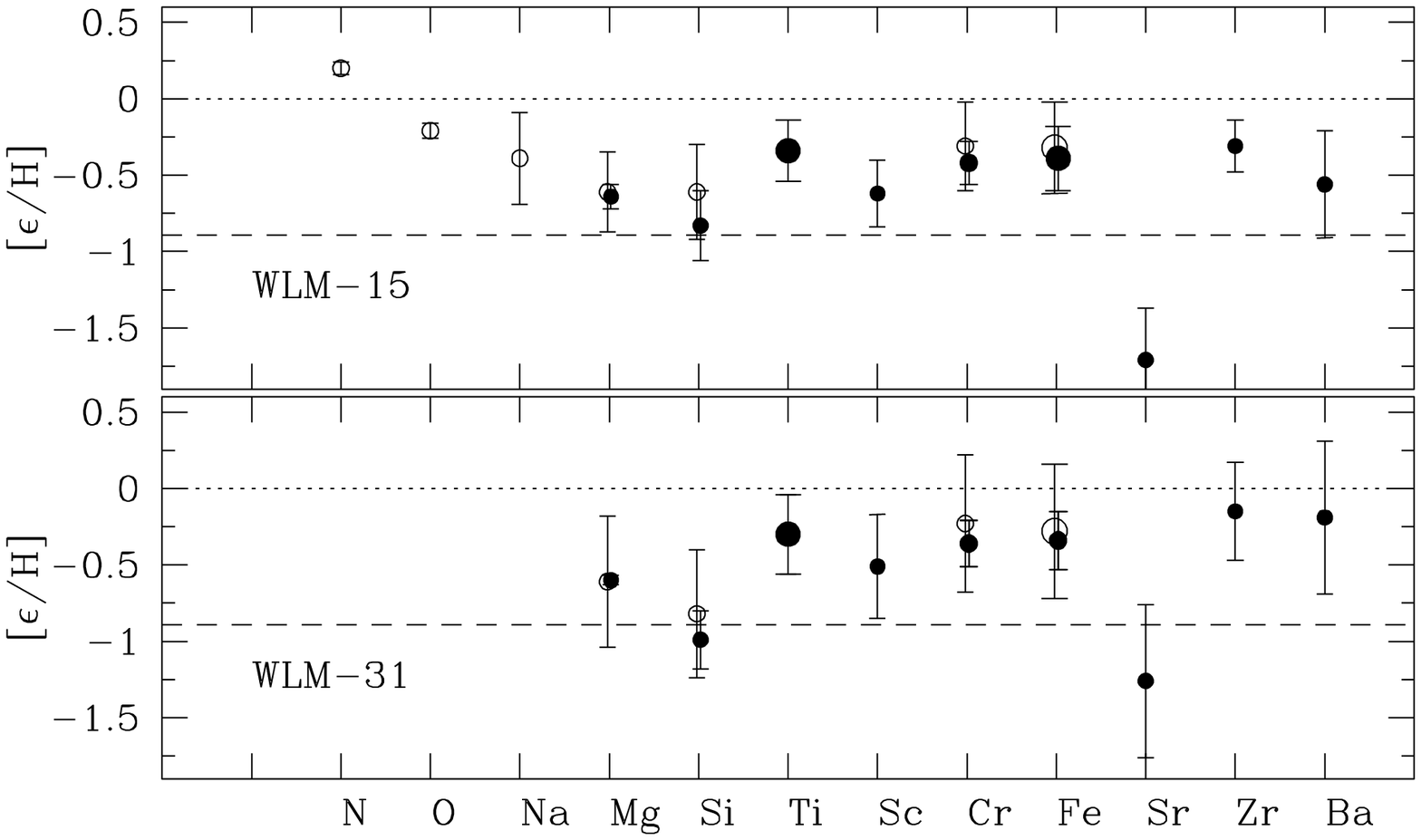}
\figcaption{
Elemental abundances for WLM-15 (top) and WLM-31 (bottom) relative
to solar (dotted line) and to the nebular oxygen underabundance
(dashed line).    Surprisingly these two stars appear more 
metal-rich than the \HII\ regions in WLM.
Two errorbars are shown for each point: thick line for the
line-to-line scatter and thin line for the systematic
uncertainties (see text and Table~\ref{atms}).
The largest data points include $\ge$30 line abundances,
and the smallest include $\le$10.
}
\label{fig5}

\clearpage
\epsscale{0.8}
\plotone{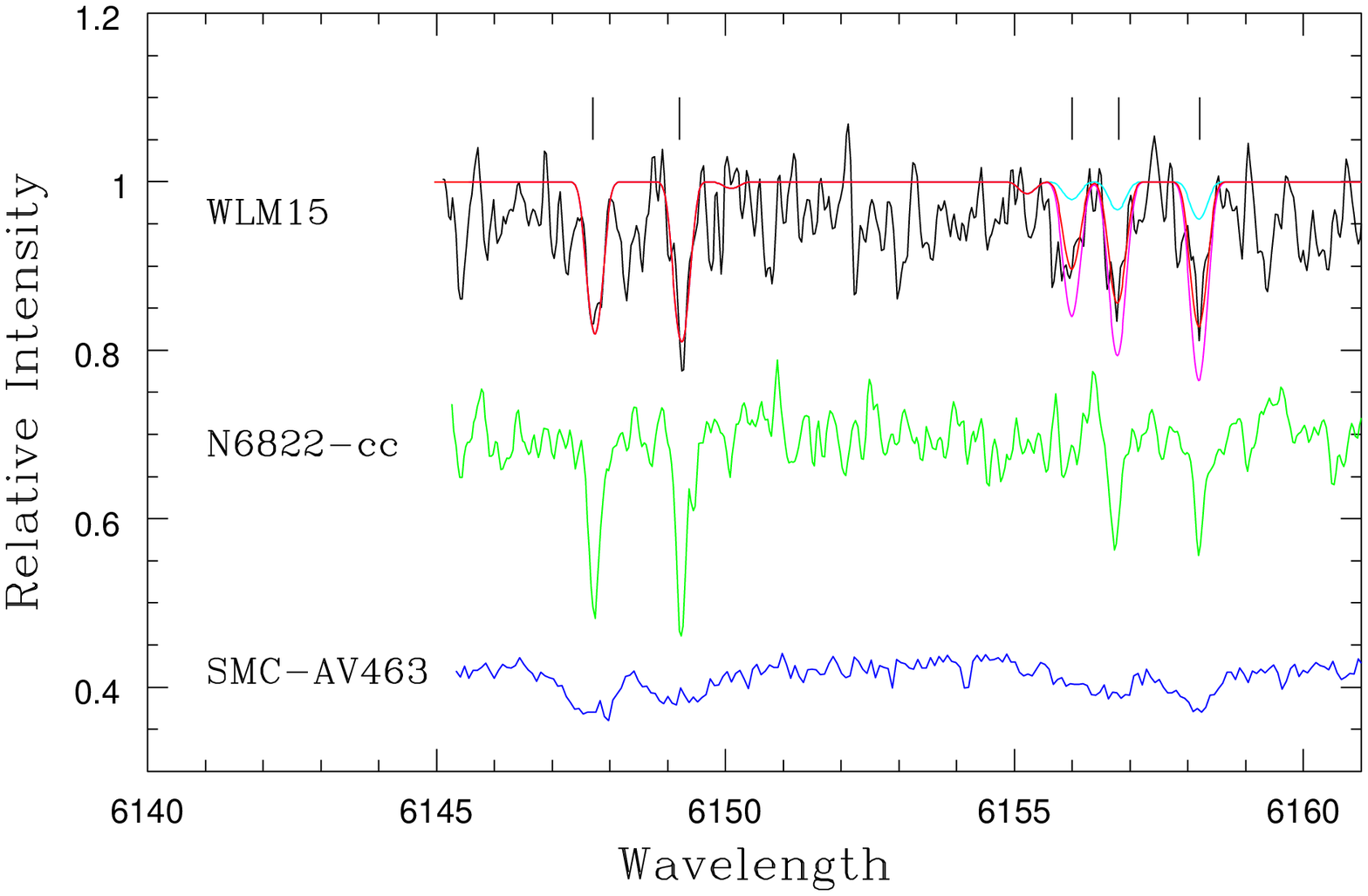}
\figcaption{
\OI 6150 spectrum synthesis for WLM-15.   Three oxygen abundances
are shown; 12+log(O/H) = 8.6 (best fit), = 8.9 (too strong),
and = 7.77 (the nebular oxygen abundance).   These are the 
LTE values; NLTE correction is $-$0.15 from Przybilla \etal (2000).
The oxygen feature is clearly well defined, 
and not reproduced by the low nebular value. 
Iron abundances for $\lambda$6147 and $\lambda$6149 line syntheses
are those from the equivalent widths analysis (Table~\ref{lines}).
Spectra of two comparison stars are shown; from model atmospheres
analysis, the oxygen abundance is the same in WLM-15 and NGC6822-cc, 
while that for SMC-AV463 is 0.2~dex lower (see text).
}
\label{fig6}

\clearpage
\epsscale{0.8}
\plotone{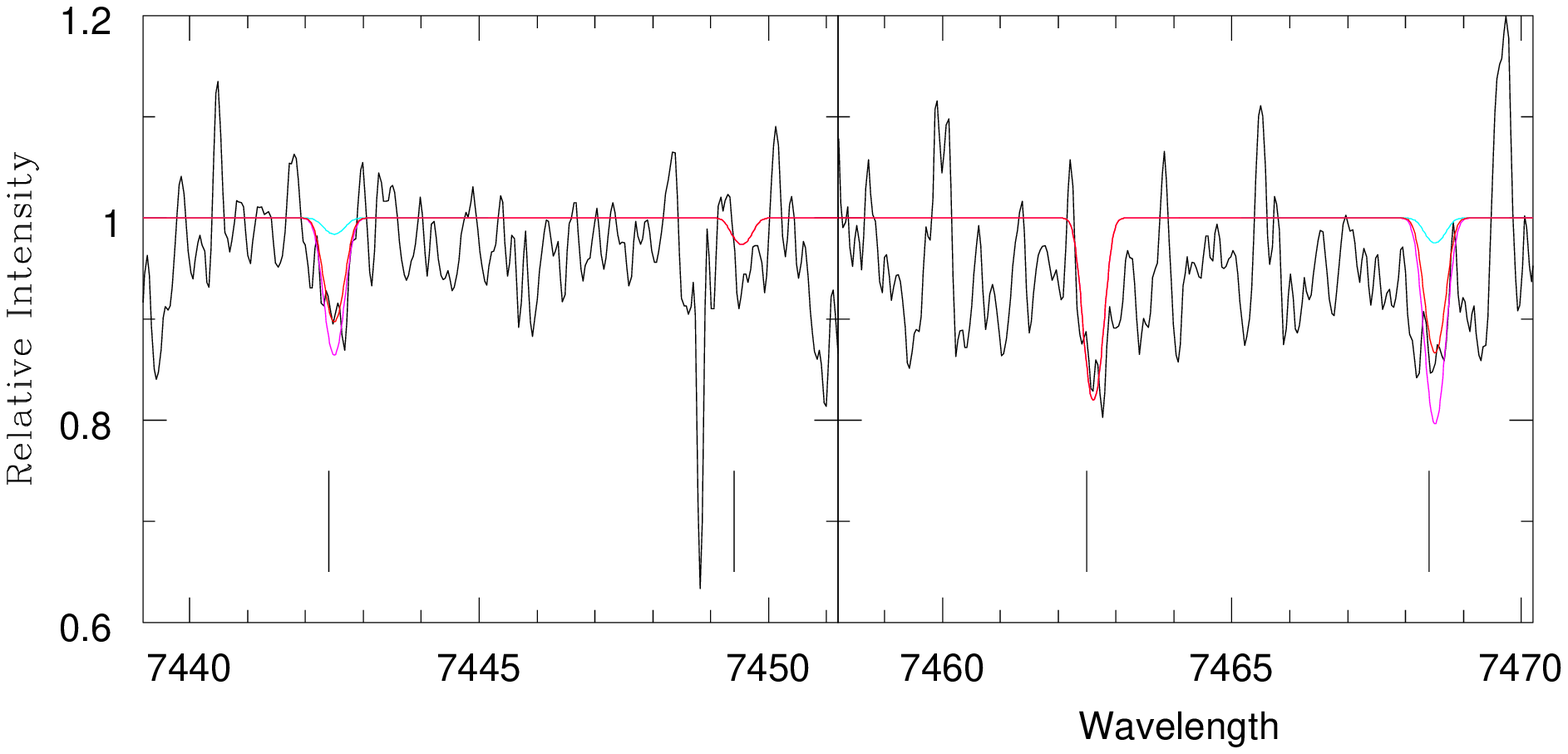}
\figcaption{
\NI\ 7440 spectrum synthesis for WLM-15.  Three nitrogen abundances are 
shown (NLTE); 12+log(N/H) = 7.5 (best fit, red line), = 7.8 (too strong,
blue line), = 6.46 (the mean nebular abundance from HM95, green line).
The iron abundance for $\lambda$7462 in the synthesis is from
the value found from its equivalent width analysis (the $\lambda$7449 
iron line is also marked, but not used). 
}
\label{fig7}

\clearpage
\epsscale{0.8}
\plotone{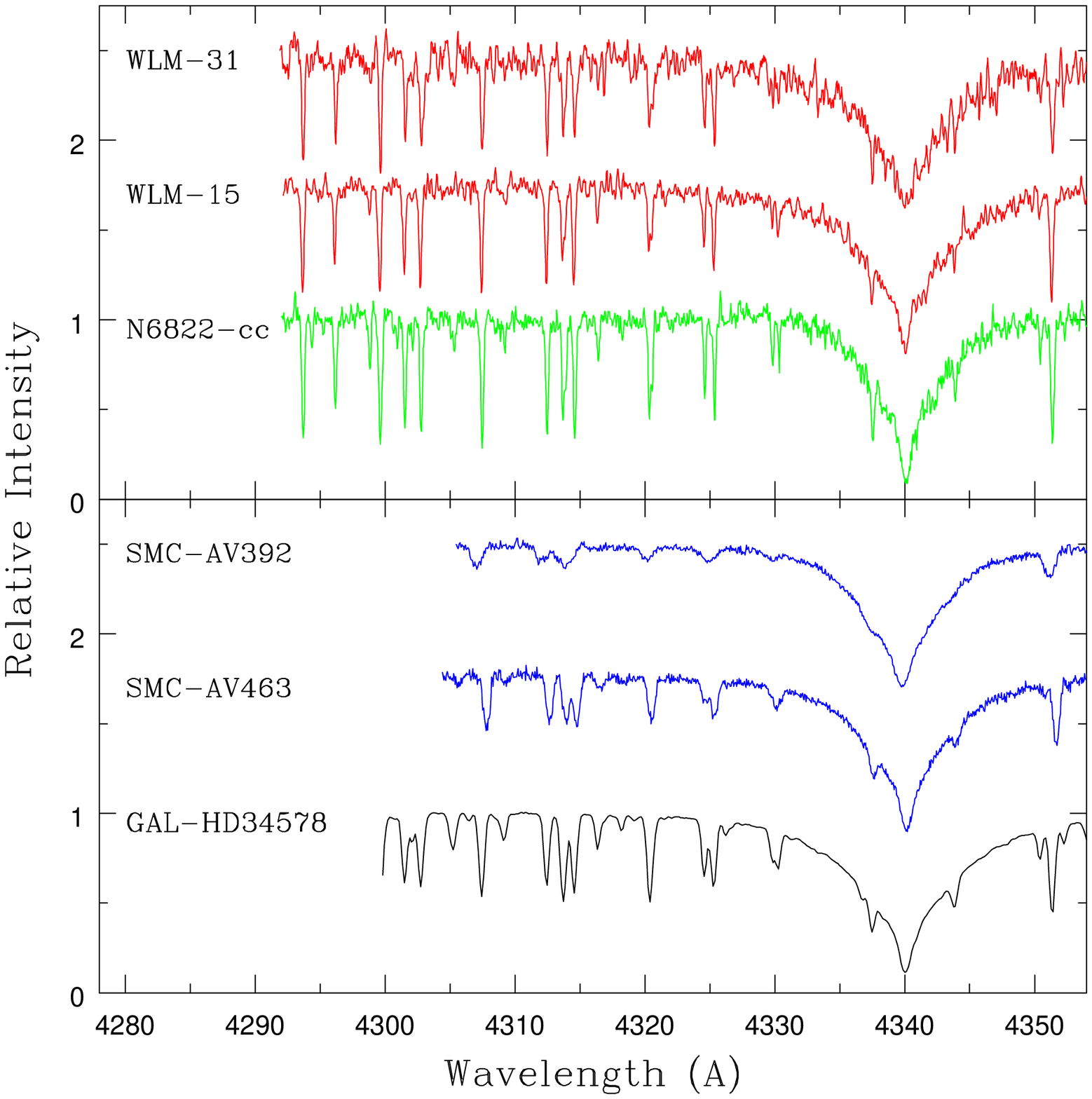}
\figcaption{
Spectra of iron-group lines around H$\gamma$ in the WLM stars and 
comparison stars with very similar atmospheric parameters 
(see Tables \ref{diff-smc} and \ref{diff-oth}).    In the top
panel, the similarities between NGC6822-cc and the two WLM stars 
are striking;  the metal lines are nearly identical.
In the lower panel, the higher broadening of the SMC and Galactic 
comparison stars make a direct comparison of the metal lines much 
more difficult.   It is obvious that the SMC stars are more metal-poor 
than the Galactic star, but detailed model atmospheres analyses are
needed to find that NGC6822-cc and the WLM stars have slightly higher
metallicities than the SMC.
}
\label{fig8}

\clearpage
\epsscale{0.8}
\plotone{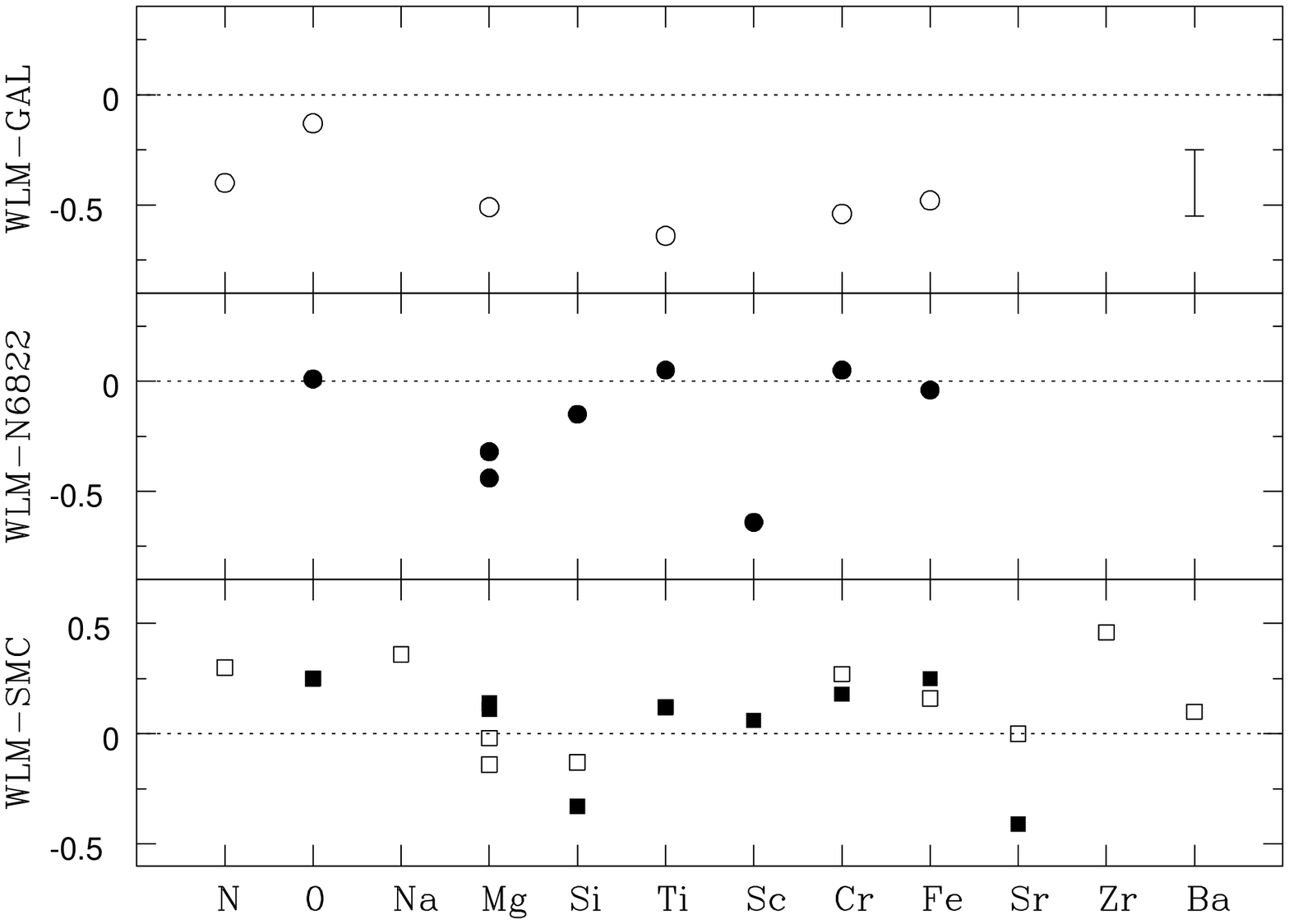}
\figcaption{
Differential abundances from Table~\ref{diff-smc} and \ref{diff-oth}, 
with the typical uncertainty shown in the top panel.  
Top panel shows that WLM-15 is more metal-poor than the Galactic
comparison star HD~34578.   Middle panel shows that WLM-15 is
comparable to NGC6822-cc, except Mg seems lower;  Sc is more 
uncertain in NGC6822-cc (see text).  Lower panel shows that 
WLM-15 is generally more metal-rich than two SMC stars
(AV392, filled squares; AV463, hollow squares),
except Mg may be lower; Si and Sr are more uncertain in 
the SMC stars themselves (see text).
The [O/Fe] ratios are similar in all stars, 
except when compared to the Galactic standard. 
}
\label{fig9}

\clearpage
\epsscale{0.8}
\plotone{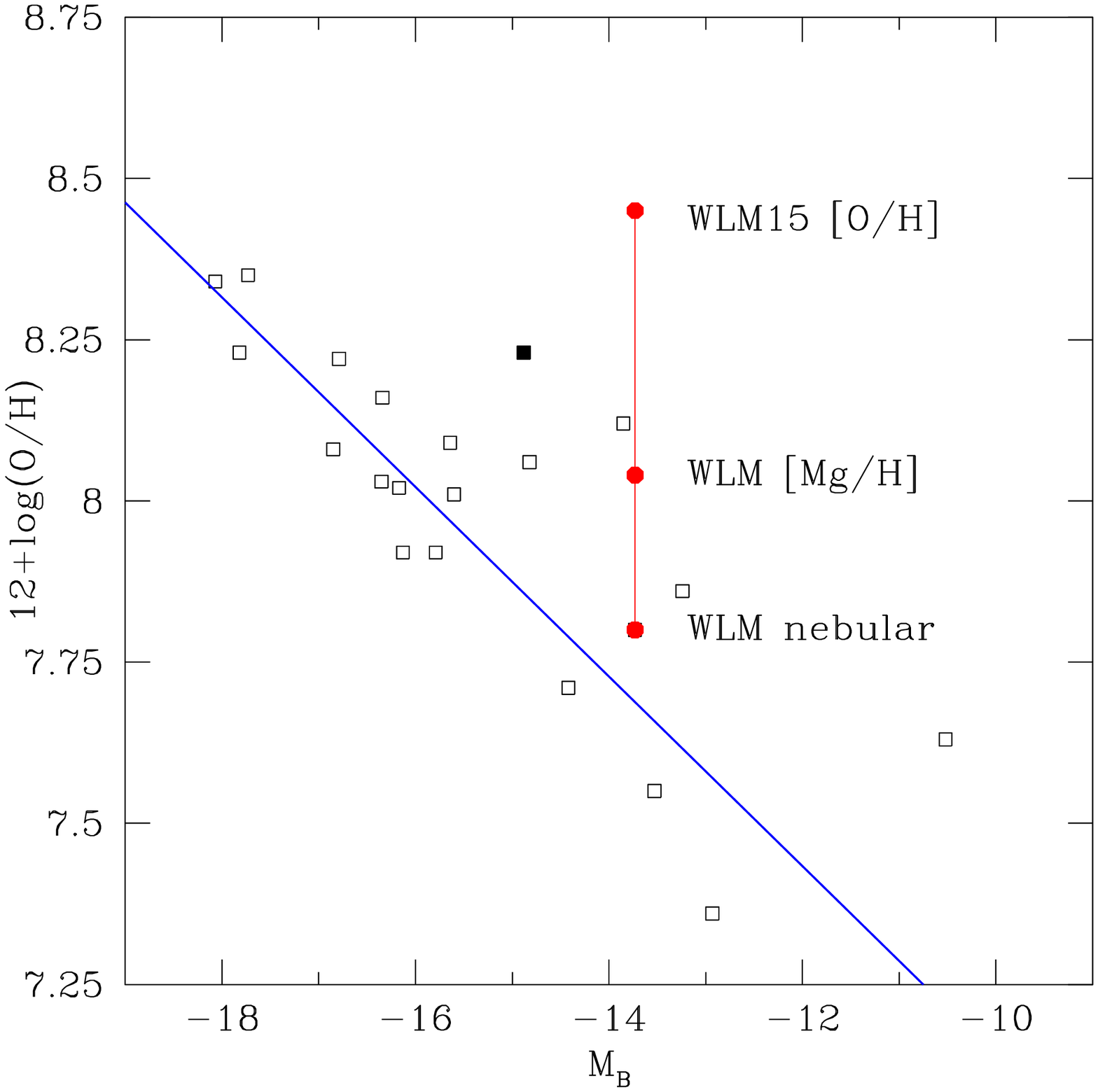}
\figcaption{
The metallicity-luminosity relationship for dwarf irregular
galaxies from Richer \& McCall (1995).    Using the same M$_B$ 
for WLM, the stellar oxygen abundance from WLM-15 is shown, as 
well as the location of the mean [Mg/H] abundance from both of
the WLM stars.   NGC~6822 is also noted ({\it solid square}).
}
\label{fig10}

\end{document}